\documentclass[iop]{emulateapj}
\usepackage{apjfonts}
\usepackage{amssymb}
\usepackage{natbib}
\usepackage{appendix}

\newcommand{\hii}{\textrm{H}~\textsc{ii}}
\newcommand{\oiii}{[\textrm{O}~\textsc{iii}]}

\newcommand{\oiiilam}{[\textrm{O}~\textsc{iii}]~\ensuremath{\lambda5007}}

\newcommand{\niilam}{[\textrm{N}~\textsc{ii}]~\ensuremath{\lambda6583}} 
\newcommand{\ha}{\ifmmode {\rm H}\alpha \else H$\alpha$\fi}
\newcommand{\hb}{\ifmmode {\rm H}\beta \else H$\beta$\fi}
\newcommand{\lya}{\ifmmode {\rm Ly}\alpha \else Ly$\alpha$\fi}
\newcommand{\dfel}{\ifmmode \Delta\log {\rm f}_{\rm EL} \else $\Delta\log$ f$_{\rm EL}$\fi}
\newcommand{\secpoint}{\mbox{$''\mskip-7.6mu.\,$}}

\def\micron{$\mu$m}
\def\msun{\ifmmode \mathrm{M}_{\odot} \else M$_{\odot}$\fi}
\def\msunyr{\ifmmode \mathrm{M}_{\odot} {\rm yr}^{-1} \else $\mathrm{M}_{\odot}$ yr$^{-1}$\fi}
\def\zsun{\ifmmode Z_{\odot} \else Z$_{\odot}$\fi}
\def\lsun{\ifmmode L_{\odot} \else L$_{\odot}$\fi}
\def\mstar{\ifmmode \mathrm{M}_{\star} \else M$_{\star}$\fi}
\def\sfrsed{SFR$_{\mathrm{SED}}$}
\def\sfruvir{SFR$_{\mathrm{UV+IR}}$}

\newcommand{\myemail}{stephane.debarros@oabo.inaf.it}

\slugcomment{}

\shorttitle{Dust attenuation of the nebular regions of $\lowercase{z}\sim2$ star-forming galaxies}
\shortauthors{De Barros et al.}

\begin{document}

\title{Dust attenuation of the nebular regions of $\lowercase{z}\sim2$ star-forming galaxies: insight from UV, IR, and emission lines}

\author{S. De Barros\altaffilmark{1, 2, 3}, N. Reddy\altaffilmark{1,4}, I. Shivaei\altaffilmark{1, 5}}

\altaffiltext{1}{Departement of Physics and Astronomy, University of California,
    Riverside, CA 92507}
    	\altaffiltext{2}{INAF-Bologna Astronomical Observatory, via Ranzani 1, I-40127 Bologna, Italy, email: \myemail}
    \altaffiltext{3}{Swiss National Science Foundation Fellow}
    \altaffiltext{4}{Alfred P. Sloan Research Fellow}
    \altaffiltext{5}{NSF Graduate Research Fellow}

\begin{abstract}
We use a sample of 149 spectroscopically confirmed UV-selected
galaxies at $z\sim 2$ to investigate the relative dust attenuation of
the stellar continuum and the nebular emission lines.  For each galaxy in
the sample, at least one rest-frame optical emission line
(\ha/\niilam\ or \oiiilam) measurement has been taken from the litterature, and 41 galaxies
have additional {\em Spitzer}/MIPS $24$\,$\mu$m observations that are
used to infer infrared luminosities.  We use a spectral energy
distribution (SED) fitting code that predicts nebular line strengths
when fitting the stellar populations of galaxies in our sample, and we perform comparisons between the predictions of our models and the observed/derived physical quantities.  We find that on average our code is
able to reproduce all the physical quantities (e.g., UV $\beta$ slopes, infrared luminosities, emission line fluxes), but we need to apply a higher dust correction to the nebular emission compared to the stellar emission for the largest SFR ($\log\mathrm{(SFR/\msunyr)}>1.82$, Salpeter IMF). We find a correlation between SFR and the difference in nebular and stellar color excesses, which could resolve the discrepant results regarding nebular dust correction at $z\sim2$ from previous results.

\end{abstract}



\section{Introduction}

\label{sec:intro}

Nebular emission (i.e.\ lines and nebular continuum from \hii\ regions) is ubiquitous in regions of massive star-formation, 
strong or dominant 
in the optical spectra of nearby
star-forming galaxies, and present in numerous types of galaxies. While 
several spectral models of galaxies including nebular emission exist
\citep[e.g.,][]{charlot&longhetti2001,fioc&rocca1997,AF03,Z08}, 
the impact of nebular emission on the determination of physical parameters of 
galaxies, in
particular at high redshift, has been largely neglected until
recently. \cite{Z08} showed that nebular
emission can significantly affect broadband photometry; the impact
becoming stronger with increasing redshift, as the equivalent width
(EW) of emission lines scales with ($1+z$). The nebular emission
contribution to broad-band photometry at high redshift can lead to
overestimates of stellar masses (\mstar) and ages
\citep[e.g.,][]{schaerer&debarros2009,onoetal2010,schaerer&debarros2010}. 

Until
recently, the derived evolution of the specific star-formation rate
(sSFR=star-formation rate/\mstar) with redshift, which indicated an
sSFR ``plateau'' at $z>2$ \citep[e.g.,][]{starketal2010,gonzalezetal2010}, was in conflict with
the expected sSFR-$z$ evolution inferred from hydrodynamical simulations
\citep[e.g.,][]{boucheetal2010,daveetal2011}, which suggest that the 
sSFR should increase with redshift.
Accounting for nebular emission can possibly reconcile the
observations with theoretical expectations of the sSFR evolution
\citep[][]{starketal2013,debarrosetal2014,duncanetal2014,salmonetal2015}. More
generally, an accurate and physically motivated modeling of nebular
emission is necessary to obtain more precise constraints on the
physical properties of high-redshift galaxies, the cosmic star-formation
history \citep{salmonetal2015}, and reionization
\citep{robertsonetal2010}.

While NIR multi-object spectrographs like MOSFIRE
\citep{mcleanetal2012} have enabled measurements of a number of
rest-optical emission lines (e.g., \hb, \ha, \oiii) up to $z\sim3.7$
\citep{holdenetal2014,kriek+15}, these lines remain
inaccessible at $z>3.7$, at least until the launch of the {\em James
  Webb Space Telescope}.  However, it is possible to find empirical
evidence of strong emission lines at $z>3.7$ by observing a flux
increase in a band which cannot be explained with pure stellar
emission and where a strong emission line is expected
\citep[e.g.,][]{charyetal2005,shimetal2011,starketal2013,debarrosetal2014,smitetal2014}. Uncertainties
remain relatively large because the inferred strength of the lines
depends on constraints on the stellar continuum and hence on other
inferred galaxy properties (e.g., age, dust extinction). One way to
avoid this issue is to model both stellar and nebular emission
simultaneously
\citep[e.g.,][]{schaerer&debarros2009,schaerer&debarros2010}. This method 
must rely on several assumptions such as the Lyman continuum
escape fraction, the metallicity, and the density and temperature of
the ionized gas.
These parameters
remain largely unconstrained at $z>3.7$, although there is mounting
evidence that ISM physical conditions evolve with redshift
\citep{kewleyetal2013b,nakajimaouchi14,steideletal2014,shapley+15}.

Until now, most studies on the impact of nebular emission on galaxy
physical properties at $z>3.7$ have assumed that both nebular and
stellar emission suffer the same amount of dust attenuation
\citep[e.g.,][]{onoetal2012,robertsonetal2010,robertsonetal2013,vanzellaetal2010a,vanzellaetal2014,labbeetal2010,labbeetal2013,oeschetal2013a,oeschetal2013b,oeschetal2014,debarrosetal2014,duncanetal2014,salmonetal2015,smitetal2014}.
However, evidence for an excess of nebular color excess relative to the
stellar color excess in local starburst galaxies has been provided
through the comparison of optical depths between Balmer lines
(\ha\ and \hb) and the stellar continuum
\citep{calzettietal1994,calzettietal1997a,calzettietal2000}. This
difference in color excess can be explained by the fact that star-formation
occurs in regions of high gas and dust column density. UV
and optical stellar continuum are produced by ionizing and
non-ionizing stars.  On the other hand, the nebular emission arises
around regions of short-lived massive stars which may not leave
their dusty birthplaces, while 
less massive stars can diffuse into less dusty regions. This could
partly explain why nebular emission shows a higher color excess than
the stellar continuum emission for local star-forming galaxies
\citep{calzettietal1997a}. The analysis presented in \cite{wildetal2011} also shows that the nebular to stellar attenuation is related to the sSFR and axial ratio.

Measuring the relative color excess of the stellar continuum and
nebular lines at high redshift has been difficult until recently \citep{reddyetal2015} due to the lack of
measurements of both \ha\ and \hb\ lines for large galaxy samples.
One way to compare the stellar and nebular color excess is to derive
the dust extinction of the stellar continuum through SED fitting and then
compare dust corrected SFRs inferred from emission lines (e.g., \ha) with
another dust-corrected SFR indicator. Under the assumption of a
constant star-formation history for $\sim100\hspace{1mm}\mathrm{Myr}$,
SFR(UV) and SFR(\ha) should equilibrate to the same value
\citep{kennicutt1998} and any remaining discrepancy between these two
SFR tracers may be due to a different dust extinction between the
regions of nebular emission and the stellar continuum, although there
are other possibilities (e.g., initial mass function variations).  The possible effect of the star-formation history must be taken into account when using SFR(\ha)/SFR(UV) to infer whether there is a difference in the color excess of the stellar continuum and nebular regions. While this kind of
comparison has been used extensively at high redshift
\citep[e.g.,][]{erbetal2006b,FS09,reddyetal2010,mancinietal2011,holdenetal2014,shivaeietal2015a},
the existence of higher nebular color excess relative to
stellar color excess remains unclear: \cite{erbetal2006b} and
\cite{reddyetal2010} find that assuming a similar color excess between the stellar continuum and nebular regions results in the best agreement between UV and Ha-based SFRs, while several other studies find that extra extinction is
required \citep{kashinoetal2013,priceetal2014}. A possible explanation for these discrepant results found at $z\sim2$ is that
the trend between the nebular versus stellar color excesses depends on
SFR, as was suggested in \cite{reddyetal2010} and
\cite{yoshikawaetal2010}, or the nebular to stellar attenuation can also depend on the sSFR \citep{wildetal2011,priceetal2014}.

Direct measurements of the Balmer decrement (\ha/\hb\ flux ratio) have been lacking for high-redshift galaxies, until recently, and these observations have
been possible because of recent new instruments as the
Wide-Field-Camera 3 of the {\it Hubble Space Telescope}, the Fiber
Multi Object Spectrograph on the Subaru Telescope
\citep{kimuraetal2010}, and the Multi-Object Spectrometer for
Infra-Red Exploration on the Keck Telescope
\citep{mcleanetal1998}. At $z\sim1.5$, \cite{dominguezetal2013},
\cite{kashinoetal2013} and \cite{priceetal2014} provide Balmer
decrement measurements for relatively large samples of galaxies, but
via stacking as the \hb\ line is typically undetected for individual
galaxies. \cite{kashinoetal2013} find a higher nebular color excess relative to stellar
color excess, but the ratio between the two is lower that the one found for local star-forming galaxies
\citep{calzettietal2000}.
The results from \cite{priceetal2014}, based on stacked data, are consistent with a difference in color excess between
nebular and stellar emission, but they also examine possible correlations between color excess ratio
and the other physical parameters (\mstar, SFR,
sSFR). 
Interestingly,
they find that the difference in the color excess between nebular and stellar emission
 increases with decreasing sSFR \citep[][Figure
  4]{priceetal2014}. They interpret this correlation with a
two-component dust model, whereby \hii\ regions are dustier than the
diffuse ISM outside stellar birth clouds. The continuum light coming
from galaxies with high sSFR would be almost completely dominated by
emission from \hii\ regions, while the emitted rest-optical continuum from low sSFR
galaxies would be dominated by less massive stars located outside
\hii\ regions. Therefore, for high sSFR galaxies, both nebular and stellar
emission would be equally attenuated and for low sSFR galaxies, nebular
emission would be more attenuated that stellar continuum \citep[][Figure
  5]{priceetal2014}. An opposite trend is found in \cite{reddyetal2015}
  with the nebular to stellar color excess ratio increasing with increasing sSFR, 
 using the largest sample including individual Balmer decrement measurements at these redshifts.

We build upon these previous studies by using the approach described in \cite{schaereretal2013}, 
  which has been already used to analyze a small sample of lensed
  galaxies in \cite{skliasetal2014}.  Namely, we derive the properties
  of 
a large number of 
individual galaxies by comparing several observables (broad-band photometry, emission line fluxes,
  infrared luminosities) with predictions from stellar population
  synthesis models that include nebular emission. We consider a range
  of star-formation histories in the SED-fitting in order to minimize the
  number of assumptions going into our analysis.  This study has
  several aims: the first is to test the ability of our SED fitting
  method to reproduce observed quantities, particularly emission line
  fluxes. Once our SED fitting code is tuned to reproduce the
  observables, we determine and compare correlations among physical
  properties of individual galaxies. Our approach is complementary to the one
  used in \cite{reddyetal2015}, because  we are using SED-fitting code to examine these issues.

The paper is structured as follows.  The selection procedure,
spectroscopic and photometric data, and the final sample used in this
work are described in Sect. \ref{sec:data}. The method used to derive
physical parameters and to model the nebular emission 
is described in Sect. \ref{sec:method}. The ability of our
method to reproduce emission line fluxes is described in
Sect. \ref{sec:results} and the implications regarding the dust extinction
toward nebular regions 
are discussed in Sect. \ref{sec:discuss}.
Section \ref{sec:conclusion}
summarizes our main conclusions.  We adopt a $\Lambda$-CDM
cosmological model with $\mathrm{H}_0=70$ km s$^{-1}$ Mpc$^{-1}$,
$\Omega_{\mathrm{m}}=0.3$ and $\Omega_{\Lambda}=0.7$. We assume a
Salpeter IMF \citep{salpeter1955}. All magnitudes are expressed in the
AB system \citep{oke&gunn1983}.


\section{Data}
\label{sec:data}

\subsection{Selection, optical and near-IR data}

All the data used in this paper are taken from the
literature. In the following, we provide only brief descriptions of
the data acquisition and reduction. The reader is invited to consult the
references given below for more details.

 \begin{figure}[htbf]
     \centering
     \includegraphics[width=8cm,trim=1.25cm 0cm 1.7cm 1.25cm,clip=true]{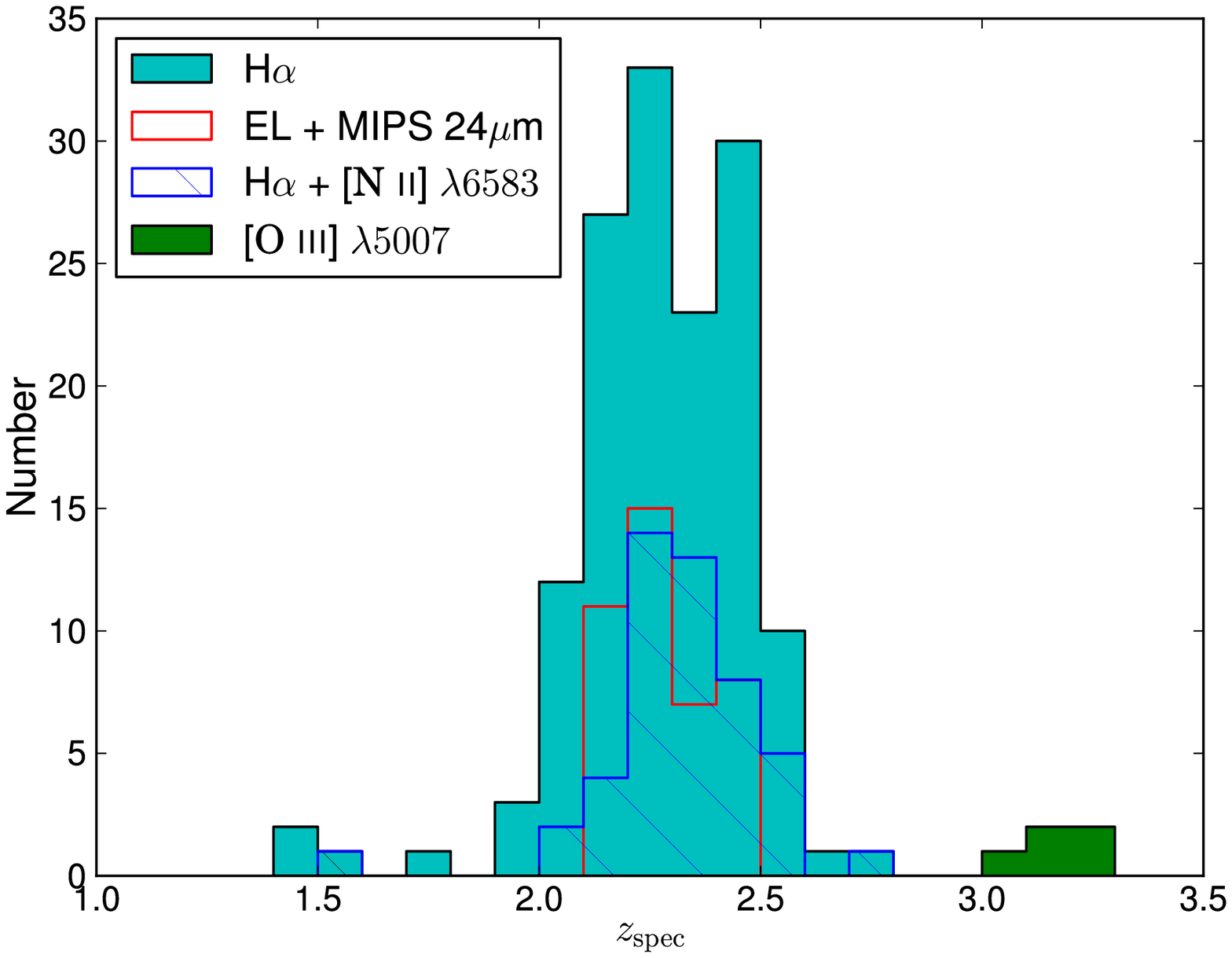}
     \caption{Redshift distribution of the star-forming galaxy sample. All galaxies have \ha\ measurements \citep{erbetal2006b,kulasetal2012,kulasetal2013}, except 5 galaxies at $z>3$ for which we have \oiiilam\ measurements \citep{kulasetal2012}. \niilam\ measurements are taken from \cite{kulasetal2013}.}
     \label{fig:histz}
 \end{figure} 
 
 We start with a large UV-selected galaxy sample, with photometric and
spectroscopic observations described in \cite{steideletal2003},
\cite{steideletal2004}, and \cite{adelbergeretal2004}. The selection
is based on $U_nG\cal R$ colors. All the galaxies in our sample are
spectroscopically confirmed with rest-frame UV observations conducted
with the Keck/LRIS spectrograph \citep{okeetal1995}.

Photometry in the $J$ and $K_S$ band are obtained with the
Palomar/WIRC and Magellan/PANIC instruments
\citep{shapleyetal2005}. These data provide important constraints on
the Balmer break, which can be used to estimate
the age of the stellar population.

Our photometry includes {\em Hubble}/WFC3 F160W observations
\citep{lawetal2012} which provide more robust constraints on the
Balmer and $4000$\,\AA\ breaks.  We combine ground-based $U_nG\cal
R$ optical photometry with $J$, $K_S$, and F160W data with {\it
  Spitzer}/IRAC observations in the four available channels
(3.6\micron, 4.5\micron, 5.8\micron, and 8.0\micron). Details
regarding the fields of view and photometry can be found in
\cite{reddyetal2006a} and \cite{reddyetal2012a}. The latter reference
provides a complete description of the data used in our study.

To more robustly constrain the physical parameters of galaxies studied in this paper, we add available near-infrared spectroscopic data taken with the NIRSPEC instrument \citep{mcleanetal1998} presented in \cite{erbetal2006c}. These observations provide \ha\ flux measurements for 91 of our galaxies. Additional NIRSPEC spectroscopy is taken from \cite{kulasetal2012}, which provides \ha\ and \oiiilam\ flux measurements for 5 and 5 galaxies respectively. We also use spectroscopic observations \citep{kulasetal2013} performed with the recently commissioned Keck/MOSFIRE near-infrared spectrometer \citep{mcleanetal2012}. This allows us to add 48 measurements of both \ha\ and \niilam\ fluxes to our study. The main source of flux measurement uncertainties for both NIRSPEC and MOSFIRE instruments come from slit losses and are estimated to be close to $\sim2$ \citep{erbetal2006b,steideletal2014}. This average  slit loss is mainly due to seeing conditions and not to the size of galaxies: for the stellar mass range probed here, at $z\sim2$, half-light radius is typically $0\secpoint2-0\secpoint3$ \citep{price+15}, well within the NIRSPEC and MOSFIRE slits used to obtain the data \citep{erbetal2006c,kulasetal2012,kulasetal2013}. Therefore we correct the observed fluxes by a factor 2. The \cite{erbetal2006c} and \cite{kulasetal2012} samples probe similar range of stellar masses and SFRs as stated in this latter reference. The galaxies from the \cite{kulasetal2013} sample are more massive on average ($10^{10}\msun\lesssim\mstar\lesssim10^{11}\msun$) but still in the stellar mass range probed in \cite{erbetal2006c} and \cite{kulasetal2012}.

Finally, combining photometric and spectroscopic data described in this section, we obtain a sample of 149 galaxies with extensive broad-band photometric coverage, and \ha\ and \oiiilam\ flux measurements for 144 and 5 galaxies respectively. \niilam\ measurements exist for a subsample of 48 galaxies with \ha\ observations.


\subsection{Mid-IR data}

MIPS 24\micron\ data in the GOODS-North field and four LBG fields
\citep[Q1549, Q1623, Q1700, and Q2343;][]{reddy&steidel2009} are used
to measure the rest-frame 8\,\micron\ fluxes of galaxies at
$z\sim2$. The complete photometry and extraction method
is described in \cite{reddyetal2010}. The measured
8\,\micron\ luminosities are then converted to the total IR luminosity
(i.e., IR luminosity integrated over 8-1000\,\micron) using
\cite{chary&elbaz2001} templates, and are corrected for the luminosity
dependance of the  conversion, as derived from stacked {\em Herschel} data for a similarly-selected sample of galaxies (Reddy et al. (2012a).
This latter reference compared L(IR) computed from 24, 100, 160\micron, and 1.4 GHz fluxes with L(IR) derived from the MIPS 24\micron\ data only: L(IR) estimates are consistent with each other, within the uncertainties of the measured IR luminosity.
We have 41
galaxies with MIPS 24\micron\ data: 21 with detections ($\mathrm{S/N}>3$) and 20 with
upper-limits. The redshift distribution of our final sample and
subsamples are shown in Figure \ref{fig:histz}.  The
constraint provided by the IR luminosity allows
 to compute the bolometric SFR (assuming energy
balance between UV and IR), and therefore the dust extinction, in an
independent way. We provide more details on how we constrain galaxy
physical parameters with IR data in Section \ref{sec:method}.

To aid our analysis, we define three different subsamples: the ``MIPS'' sample (galaxies with MIPS data), the ``MIPS detected'' sample (galaxies detected at 24\micron\ with $>3$ sigma), and the ``MIPS upper limit'' sample (galaxies undetected), with 41, 21, and 20 galaxies respectively.  All these galaxies have at least \ha\ measurements, with 12 having \niilam\ measurements.

 \begin{figure*}[htbf]
     \centering
     \includegraphics[width=18cm,trim=1.25cm 4cm 1.75cm 4.5cm,clip=true]{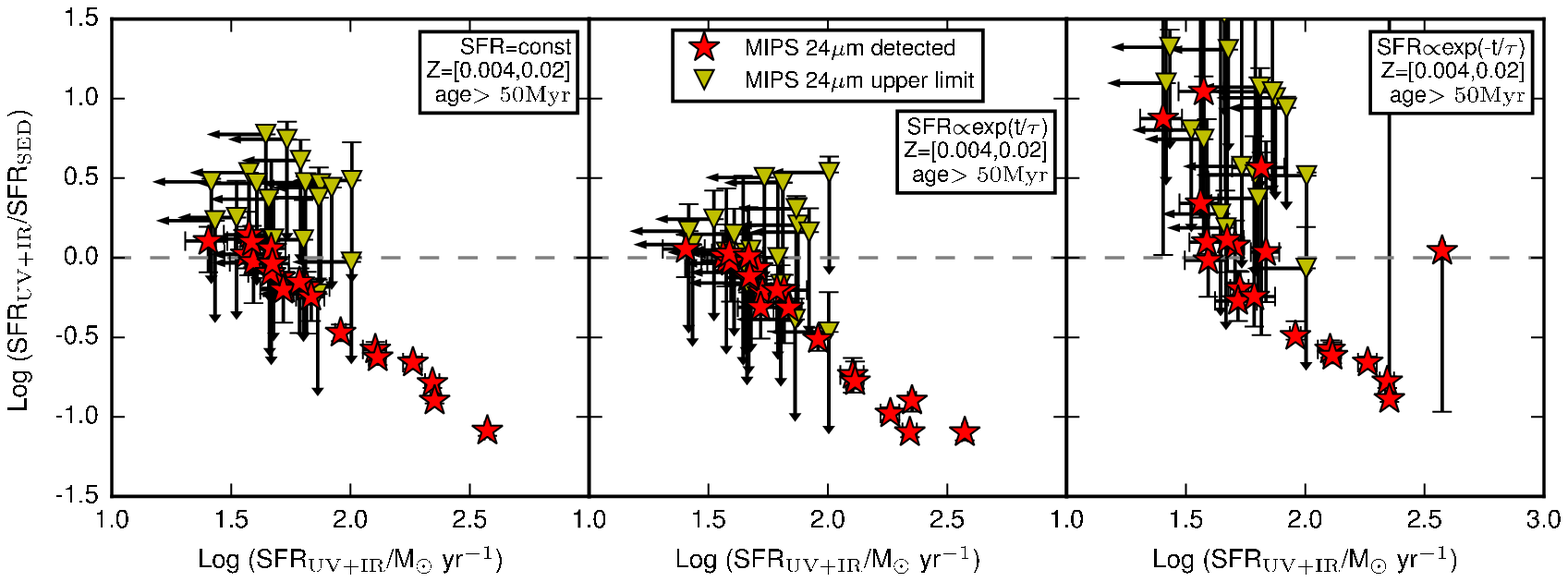}
     \caption{Comparison between log \sfruvir\ and log(\sfruvir/\sfrsed) for three different SFHs (constant, rising, and declining, from left to right). An increasing inconsistency between \sfruvir\ and \sfrsed\ appears at \sfruvir$\ge10^{1.95}$ \msun yr$^{-1}$ for galaxies detected in MIPS 24\micron. We remind that we assume energy balance in our SED fitting procedure.}
     \label{fig:uvir_sed}
 \end{figure*} 
 
 
\section{SED fitting}
\label{sec:method}

Physical properties of the galaxies are inferred by fitting model
SEDs 
to the broad-band photometry from $U$-band to 8\micron.
Our SED fitting code is a modified version of the photometric redshift
code {\it HyperZ} \citep{bolzonellaetal2000}. We generate a set of
spectral templates with the GALAXEV code of \cite{BC03}, for three
different metallicities (Z=0.004, 0.04, 0.02) and three different
star-formation histories (SFH=constant, exponentially rising,
exponentially decreasing). The timescales used for the rising and
declining SFHs are respectively $\tau_r=[10, 30, 50, 70, 100, 300,
  500, 700, 1000, 3000]$ Myr and $\tau_d=[10, 30, 50, 70, 100, 300,
  500, 700, 1000, 3000, \infty]$ Myr. The stellar age is defined as
the age since the onset of star-formation. Unless stated otherwise, we
consider a minimum age of 50 Myr corresponding to the typical 
dynamical timescale expected for $z\sim2$ galaxies \citep[e.g.,][]{reddyetal2012a}.

We assume that the dust attenuation is described by the Calzetti dust attenuation
curve \citep{calzettietal2000}, with 
$A_V$ allowed to vary from 0.0 to 6.0, in steps of 0.05. The
intergalactic medium (IGM) is treated using the Madau prescription
\citep{madau1995}. Redshifts are fixed to the spectroscopic ones.  For
our entire grid of models, we compute the $\chi^2$ and the scaling
factor of the template; the latter determines the SFR and \mstar.

We account for the effect of nebular emission (both continuum and
emission lines) following \cite{schaerer&debarros2009} and
\cite{schaerer&debarros2010}. The strength of nebular emission depends
mainly on the number of Lyman continuum photons, which is computed
from stellar population synthesis models. Relative line intensities
are taken from \cite{AF03} and \cite{storey&hummer1995}, for typical
ISM conditions ($n_e=100$ cm$^{-3}$, $\mathrm{T}=10^4$ K). We assume
that the Lyman continuum escape fraction is equal to zero, which means
that our models produce the maximum theoretical strength for nebular
emission for the adopted ISM physical conditions. 
Additionally, we assume that the same dust attenuation curve
applies to the nebular and stellar emission, and that the color excess
derived from the continuum applies to the line emission. If, as in the
local universe, the nebular lines are subject to larger color
excess than UV continuum, then our assumption would over-predict the emission line
fluxes compared to the measured fluxes. \cite{reddyetal2010} show that the impact of nebular emission on ages and stellar masses at $z\sim2$ is insignificant because most of the longer wavelength bands (e.g., IRAC) used in the SED fitting are unaffected by line emission at these redshifts. Therefore our assumptions about
nebular emission modeling, particularly about nebular attenuation,
does not significantly affect the physical parameters derived from the
SED modeling.

We also ignore the \lya\ line
contributions to the broadband fluxes because over 122 galaxies
with \lya\ flux measurements available in our sample, only one third (44)
exhibits \lya\ in emission, and more than a half of this third
have EW(\lya)$<10$ \AA. Therefore we set the \lya\ flux to
zero in our nebular emission modeling.

We used the IR luminosity as
an additional constraint to the SED fitting by assuming a balance
between the energy absorbed in the UV/optical (we integrate over 912\AA\ to 3\micron), and re-radiated in the
IR.  {\it HyperZ} builds
a grid of models (Hypercube). For each model in our grid, we compute the
expected IR luminosity (8-1000\micron) and then exclude from the grid
all the solutions for which L(IR)$_{\mathrm{SED}}$ is inconsistent
with L(IR)$_{24\mu\mathrm{m}}$ within 68\% confidence.
For the non detections, we exclude solutions inconsistent with the
L(IR)$_{24\mu\mathrm{m}}$ 3$\sigma$ upper limits.  The 41 galaxies in the MIPS sample
will have the most robust determinations of SFRs and dust reddening
given the independent information
provided by the IR constraints. We have inspected the SED for each object
to ensure that the fits are still good when assuming the energy balance: there is no significant difference
between with and without energy balance (comparison based on $\chi^2$), except for a small subsample (see Section~\ref{sec:compsfr}).

Minimization of $\chi^2$ over the entire parameter space yields the
best-fit parameters. Uncertainties are determined from the $\chi^2$
distribution, through the likelihood marginalization for each
parameter of interest with $\cal{L}\propto\exp$$(-\chi^2/2)$.


\section{Results}
\label{sec:results}

In this section, we first 
compare the SFRs derived through SED fitting
with \sfruvir\ (Section~\ref{sec:compsfr}) and 
determine if the assumed Calzetti reddening curve is appropriate (Section~\ref{sec:attlaw}).
Because our final sample has been obtained after 
applying several criteria (UV-selected, optical emission line measurements), we
compare the general properties of our sample with previous studies to
ensure that it is representative of $z\sim2$ star-forming galaxies (Section~\ref{sec:genprop}).
We then test the ability of 
the SED models to reproduce observed emission line fluxes for the MIPS detected sample (Section~\ref{sec:reprodel}) and explore if there is any trend among the physical parameters that could explain the discrepant results regarding the 
difference in color excess between nebular lines and stellar continuum found in previous studies (Section~\ref{sec:corr}).


\begin{figure}[htbf]
\centering
\includegraphics[width=9cm,trim=0.5cm 0.2cm 0.75cm 0.75cm,clip=true]{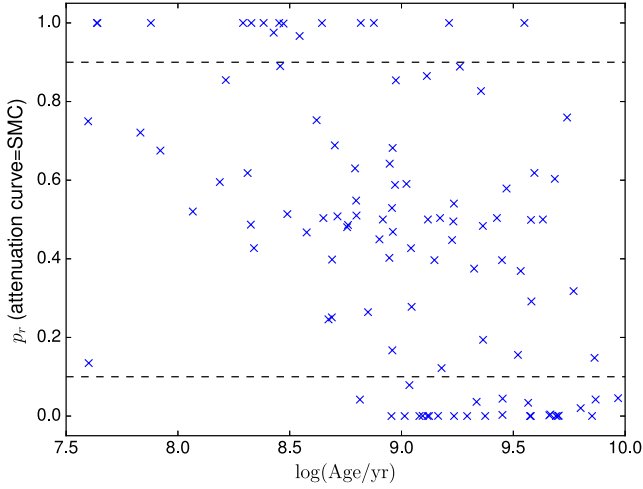}
\caption{Relative probability for the SMC 
extinction curve to be the best-fit
  dust extinction curve versus age (also derived assuming an SMC curve). We
  show the result assuming a constant SFH for the entire sample. The
  dashed lines show the threshold we adopt when determining whether a model 
is the best fit one
($p_r\geq0.9$). All galaxies above the upper line are
  considered best fit with the SMC curve and all galaxies below the
  lower line are considered best fit with the Calzetti curve. We
  introduce a slight offset in age to make the figure clearer.}
\label{fig:prsmc}
\end{figure} 

\subsection{\sfrsed\ versus \sfruvir}
\label{sec:compsfr}

Figure \ref{fig:uvir_sed} shows the comparison of the SFRs derived
from SED fitting with those computed from the sum of the obscured (IR)
and unobscured (UV) SFRs (the latter assume the
\citealt{kennicutt1998} relations).  We find a good agreement between 
\sfrsed\ and \sfruvir\ for \sfruvir$\lesssim10^2\msunyr$ for the rising and constant SFHs: for the MIPS detected sample, the difference is $\lesssim0.25$ dex. There is a mismatch between
\sfrsed\ and \sfruvir\ for 7 galaxies which are the only known ultra-luminous galaxies 
(ULIRGs, $\mathrm{L(IR)}>10^{12}\lsun$) in our
sample. For these galaxies, the discrepancy between \sfrsed\ and \sfruvir\ is $>0.5$ dex, and can be as large as $\sim1$ dex.
We further investigate the ULIRGs physical properties in 
Appendix~\ref{sec:ulirg}. The declining SFH leads to lower \sfrsed\ in comparison with \sfruvir. This is consistent with the fact that declining SFHs can underpredict SFRs \citep[e.g.,][]{reddyetal2010,wuytsetal2011,reddyetal2012a,priceetal2014}. In the following, we will show and discuss the results for the rising SFH mainly, since the constant and rising SFHs lead to similar results (except as indicated otherwise).

In the remainder, we present results based on SED fitting
assuming energy balance, but we exclude these 7 ULIRGs from our analysis.


\subsection{Attenuation curve}
\label{sec:attlaw}

While we assume a Calzetti attenuation curve to perform SED fitting,
\cite{reddyetal2010} show that the IRX-$\beta$ relation is more
consistent with an SMC-like curve
\citep{prevotetal1984,bouchetetal1985} than the Calzetti curve for
galaxies younger than 100 Myr.
To check if this is also true for our sample, we use the
Akaike Information Criterion \citep[AIC; ][]{A74} which allows us to compare different models and
derive the relative probability for one model to be the best model among a given set of models.
In our case, we compare modeling assuming the Calzetti curve with those that assume the SMC curve (where all other SED parameters have been fixed in the fitting).
AIC is defined
as:
\begin{equation}
\mathrm{AIC}=\chi^2+2 k
\label{eq:aic}
\end{equation}
where $k$ is the number of free parameters. From the AIC, we derive the relative probability $p_r$ for the $i$th model (model assuming SMC/Calzetti attenuation curve) to be the best model as:
\begin{equation}
\Delta_i=\mathrm{AIC}_i-\mathrm{min(AIC)}
\label{eq:Daic}
\end{equation}
\begin{equation}
p_r=\frac{\exp(-1/2\Delta_i)}{\Sigma_i\exp(1/2\Delta_i)}
\label{eq:paic}
\end{equation}
\begin{figure}[htbf]
\centering
\includegraphics[width=9cm,trim=0.5cm 0.2cm 0.75cm 0.75cm,clip=true]{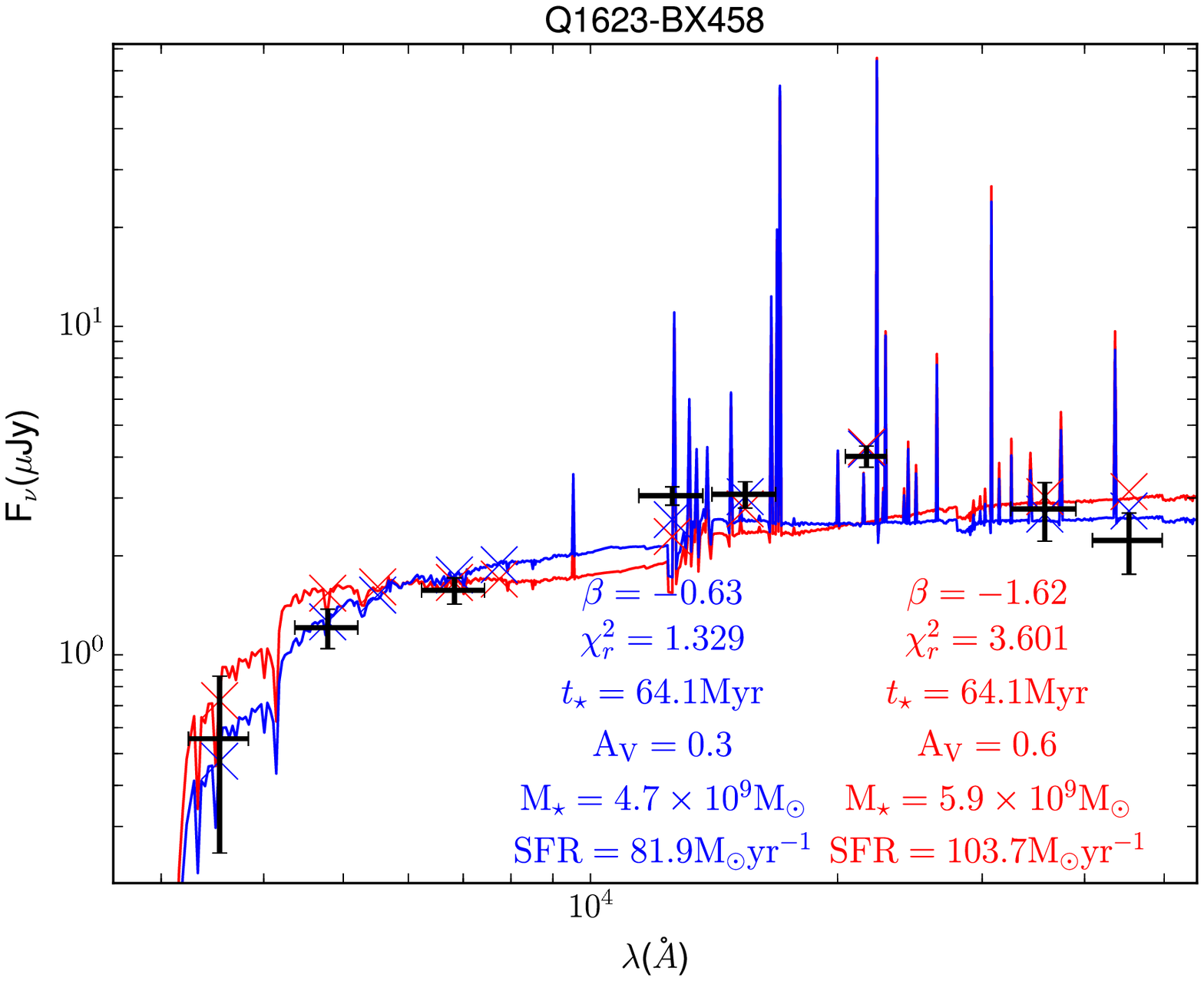}
\includegraphics[width=9cm,trim=0.5cm 0.2cm 0.75cm 0.75cm,clip=true]{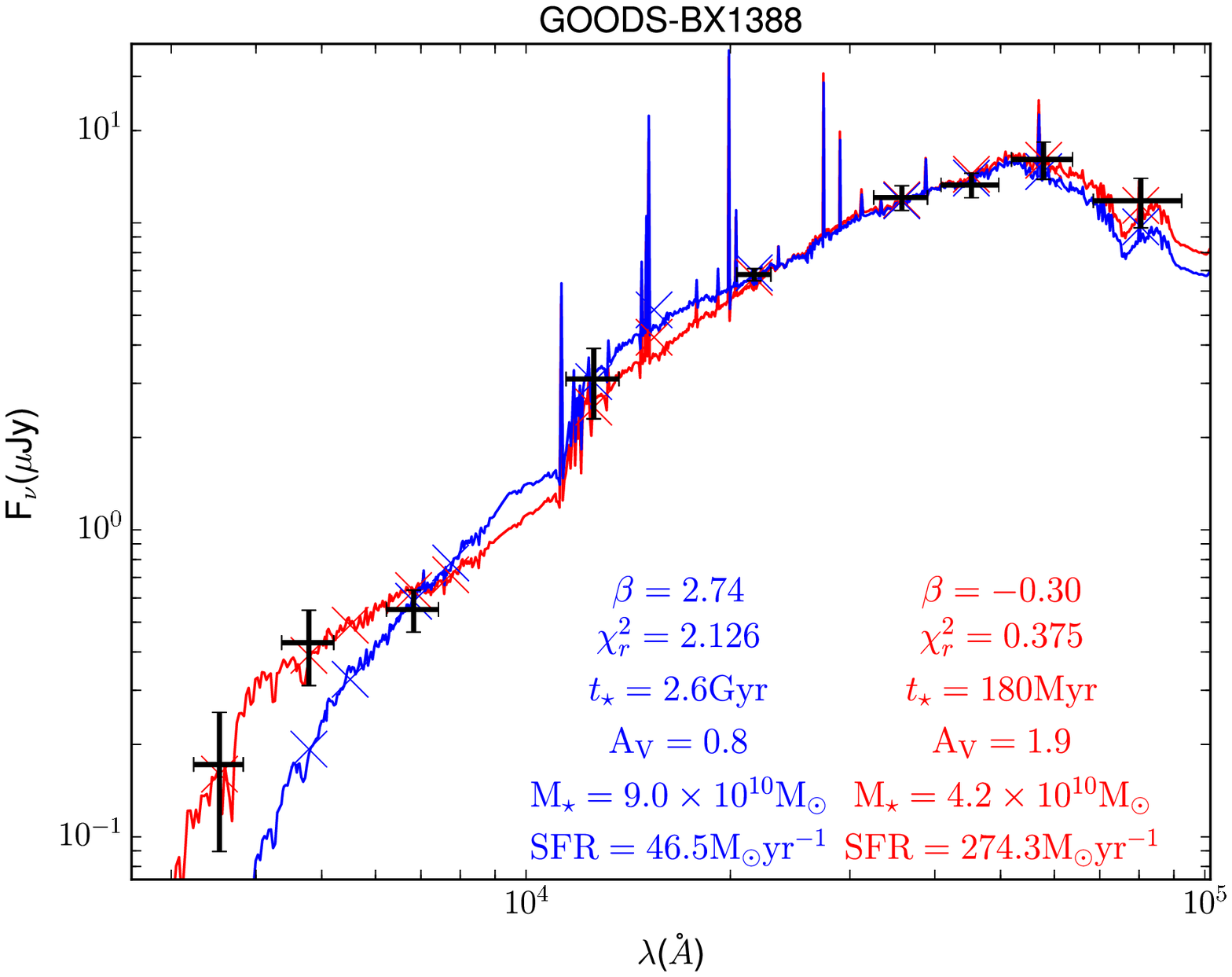}
\caption{Observed SEDs (thick black crosses) and best-fit assuming a constant SFH with a Calzetti attenuation curve (in red) and a SMC attenuation curve (in blue) for a galaxy best fit (i.e.~$p_r>0.9$) with an SMC curve (top) and a galaxy best fit with the Calzetti curve (bottom). Thin red and blue crosses show integrated fluxes in each band. The main physical properties are shown in blue and red assuming the SMC and the Calzetti curves respectively. Apart from the use of SED fitting to discriminate among attenuation curves, it is worth noting that the $K_s$ band excess for the galaxy Q1623-BX458 can only be explained by \ha\ emission \citep{shivaeietal2015a}.}
\label{fig:sedsprob}
\end{figure} 
We show the results obtained for the entire sample (Figure
\ref{fig:prsmc}). 
Since AIC is only a statistical test, we define a
threshold where we consider a model as the best fit one if the
relative
probability is $p_r\geq0.9$. While galaxies best fit with the SMC curve are found at any age, galaxies of a young age ($<10^{8.5}$ years) are more likely to be fit with a SMC-like attenuation curve, while galaxies best fit with the Calzetti curve are $\gtrsim10^9$ years. This relation between age and attenuation curve is similar to previous results in literature based on the comparison between UV to IR luminosity ratio and $\beta$ slope \citep{reddyetal2006b,reddyetal2010,reddyetal2012b,sianaetal2008,sianaetal2009,wuyts+12,shivaeietal2015a}.

 \begin{figure*}[htbf]
     \centering
     \includegraphics[width=18cm,trim=1.5cm 4cm 1.75cm 4.5cm,clip=true]{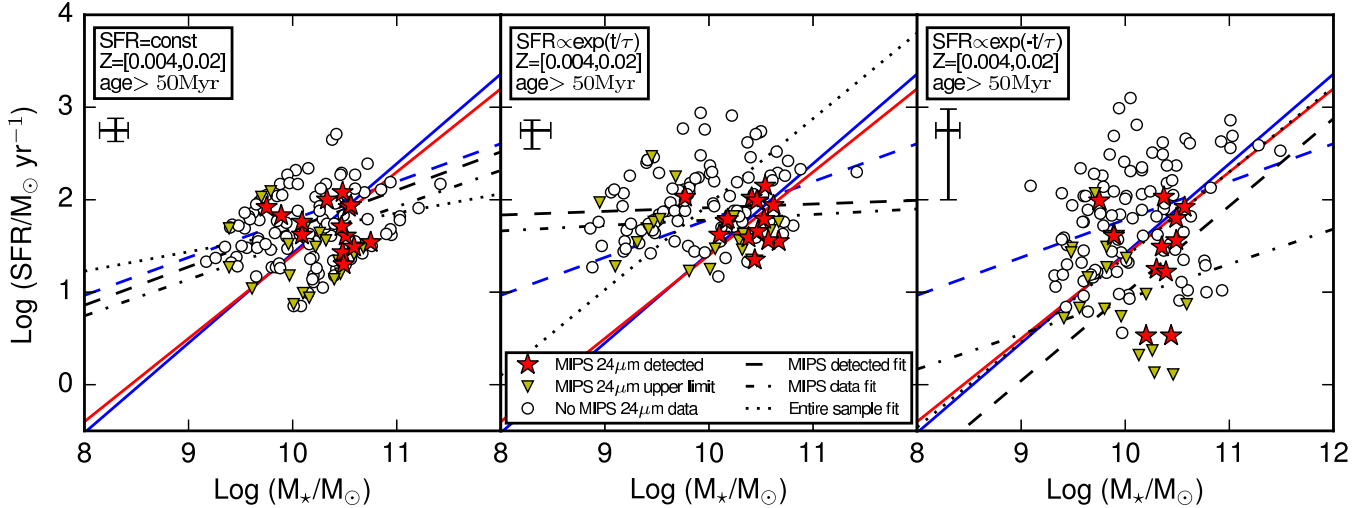}
     \caption{Star-formation rate versus stellar mass for three
       different star-formation histories. Star-formation rates are
       inferred from SED fitting assuming energy balance when MIPS
       24\micron\ data are available (see Section
       \ref{sec:method}). The red and blue lines show the star formation
       main sequence from \cite{daddietal2007} and
       \cite{reddyetal2012a}, respectively, at $z\sim 2$. The blue dashed line shows a linear fit to the data from \cite{reddyetal2012a}.  We show linear fits to our
       data for three different samples: MIPS detected (dashed lines),
       MIPS data (dash-dot lines) and the entire sample (dotted
       lines). Typical error bars are shown in the top left corner for
       each SFH.}
     \label{fig:mstarsfr}
 \end{figure*} 

       \begin{figure}[htbf]
     \centering
     \includegraphics[width=8cm,trim=0.5cm 0cm 1.7cm 1.25cm,clip=true]{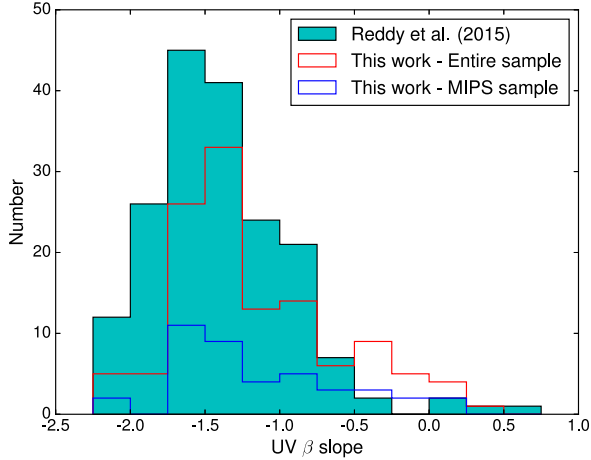}
     \caption{UV $\beta$ slope distributions for the \cite{reddyetal2015} sample, our entire sample, and the MIPS sample. The $\beta$ slopes are derived from the observed SEDs.}
     \label{fig:histo_beta}
     \end{figure} 

We show in Figure~\ref{fig:sedsprob} two examples: one galaxy best fit with the SMC attenuation curve and one best fit with the Calzetti curve ($p_r(\mathrm{SMC})>0.99$ and $p_r(\mathrm{SMC})<0.01$, respectively). The main difference between the solutions obtained assuming different attenuation curves is the UV $\beta$ slope ($f_\lambda\propto\lambda^\beta$). The UV $\beta$ slope  depends on the age and dust attenuation of the galaxy \citep[and the SFH,][]{leitherer&heckman1995,meureretal1999}.
In the two examples shown, Balmer breaks are constrained by the $J-\cal R$ color but in each case, only one attenuation curve is able to reproduce both the Balmer breaks and the UV $\beta$ slopes. Best fit assuming the Calzetti curve leads to $\beta$ slope bluer than the slope derived with the SMC curve. For young galaxies ($<100\mathrm{Myr}$), the Calzetti curve leads to $\beta$ slopes too blue compared to the observed ones (Figure~\ref{fig:sedsprob} top), while for old galaxies, the SMC curve leads to $\beta$ slopes too red (Figure~\ref{fig:sedsprob} bottom). Despite the well known degeneracy between dust attenuation and age derivation, the $J-\cal R$ color is enough to lift the degeneracy for very young and very old galaxies, allowing to statistically discriminate between the SMC and the Calzetti curve.

According to our statistical test, the percentage of galaxies best
fit with the SMC curve is $<10\%$ of the MIPS sample (for any SFH), 
while galaxies
best fit with the Calzetti curve represent $\sim25\%$. The rest of the
sample does not have relative probability passing our threshold, and
it represents $\sim65\%$ of our sample. If we consider probabilities at face values, $\sim20\%$
of the galaxies are best fit with an SMC curve and $\sim80\%$ are best fit with a Calzetti curve. Because the majority of galaxies in our sample appears to be better described with the
Calzetti curve, we assume this attenuation curve in the rest of
Section~\ref{sec:results}.


\subsection{General properties of the sample}
\label{sec:genprop}

In this section, we compare the properties of the MIPS sample to 
those of the entire sample and with the properties of $z\sim2$
star-forming galaxies taken from the literature in order to highlight
the relevance of our method and the representativeness of our sample.

The stellar masses and SFRS derived assuming a constant SFH are consistent with the relation between \mstar\ and SFR derived in
\cite{daddietal2007} and \cite{reddyetal2012a}, with a rms scatter of 0.37 dex as in \cite{reddyetal2012a}
(Figure \ref{fig:mstarsfr}).  The rms scatter for the rising and the declining SFHs are 0.56 and 0.75 dex, respectively.
Using a Spearman correlation test, we find that SFR and \mstar\ are weakly correlated (the strongest correlation is found for the declining SFH, with $\rho=0.22$ and $\sigma=2.56$).
We perform linear fits to our sample and subsamples (total MIPS sample and the MIPS-detected sample). While the stellar masses and SFRs for the MIPS detected sample do not depend strongly on the assumed star formation history, for galaxies without MIPS data the effect of star-formation history is more important. While ages derived with the rising star formation histories are similar to those derived assuming a constant SFH ($\overline{\Delta(\log\mathrm{Age)}}=0.0$, $\sigma=0.2$), dust extinctions are higher relative to the ones derived with the constant SFH ($\overline{\Delta(\log\mathrm{A_\mathrm{V}})}=0.2$, $\sigma=0.2$). Accordingly the SFRs are also higher \citep[$\overline{\Delta(\log\mathrm{SFR}})=0.3$, $\sigma=0.3$, see also][]{SP05,debarrosetal2014}. For objects in the MIPS sample, the differences in SED parameters obtained with different SFHs are reduced relative to those derived when no IR constraints are available. 

       \begin{figure}[htbf]
     \centering
     \includegraphics[width=8cm,trim=0.5cm 0cm 1.45cm 1.25cm,clip=true]{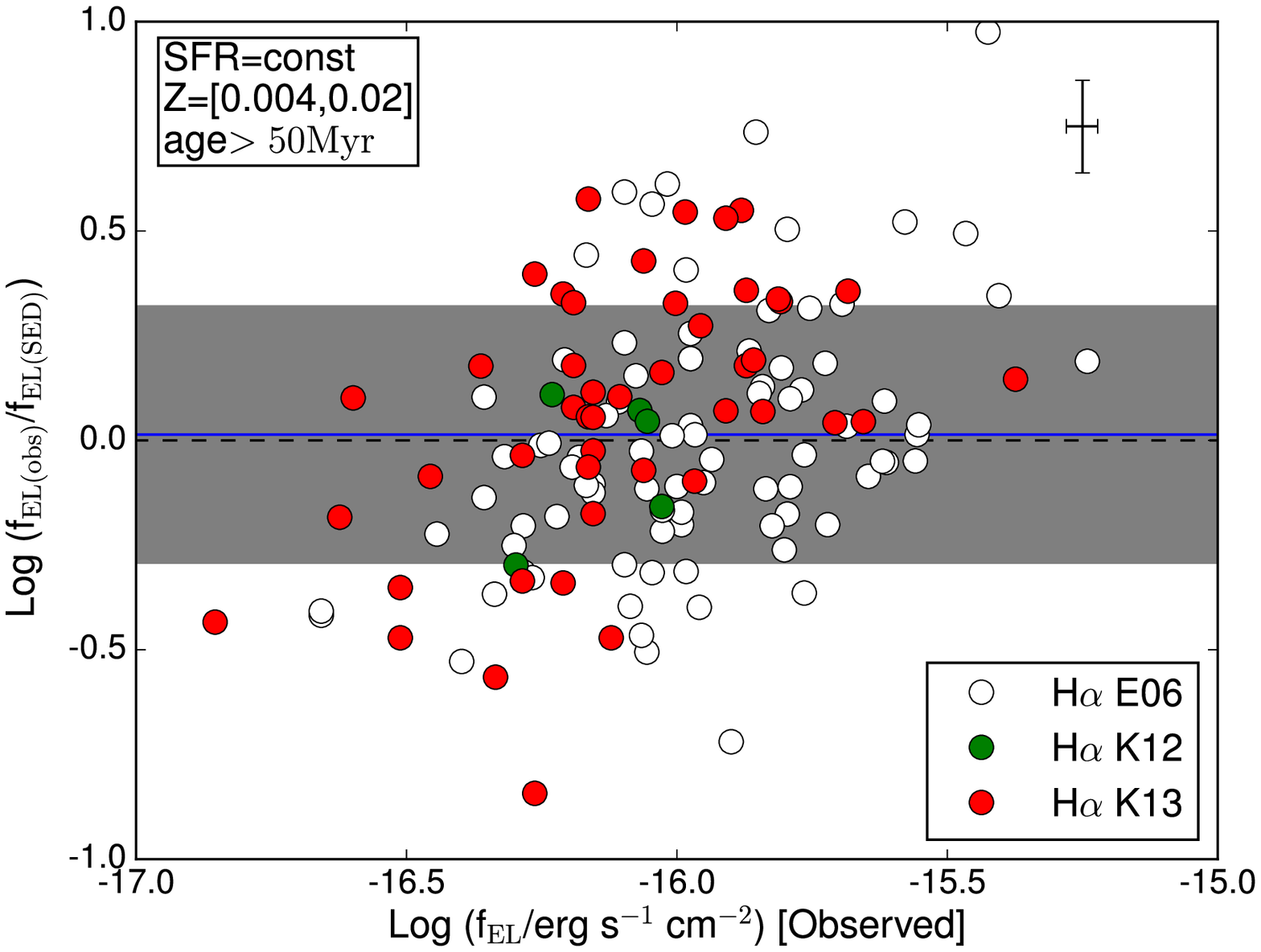}
     \caption{Comparison between $\log\mathrm{f}_{\mathrm{EL(obs)}}$ and $\log(\mathrm{f}_\mathrm{EL(obs)}/\mathrm{f}_\mathrm{EL(SED)})$ for the entire sample. Emission line fluxes are taken from \cite{erbetal2006c}, \cite{kulasetal2012}, and \cite{kulasetal2013}, hereafter E06, K12, and K13 respectively. The blue line shows the median $\log(\mathrm{f}_\mathrm{EL(obs)}/\mathrm{f}_\mathrm{EL(SED)})$  value and the grey area the 1$\sigma$ dispersion.} Errors are determined by combining measurement error of the observed fluxes with the errors derived from parameter degeneracy (black cross in the right top corner).
     \label{fig:Del_mips}
     \end{figure} 

While deriving the relation between SFR and \mstar\ at $z\sim2$ is 
not the main focus of this paper, our method used to derive physical
properties of galaxies leads to results consistent with what we
 expect for typical $z\sim2$ star-forming galaxies in terms of their
SFRs and stellar masses, if we assume a constant SFH.

To assess if our sample is biased in terms of dust properties, we compare observed UV $\beta$ slopes from the sample from \cite{reddyetal2015} with the slopes of our sample in Figure~\ref{fig:histo_beta}. We derive the slopes using the filters probing the rest-frame range $\lambda=1500-2500$\AA\ \citep[e.g.,][]{castellanoetal2012}. Both our entire and MIPS samples exhibit distributions similar to the \cite{reddyetal2015} sample, with a small galaxy fraction probing redder $\beta$ slopes ($\beta>-0.5$).

Thus, our sample is likely representative of  $>10^9\msun$ $z\sim2$ galaxies in general.
 
 
 \subsection{Reproducing emission line fluxes}
 \label{sec:reprodel}
 
We now compare the emission line fluxes predicted by our SED fitting
code (see Section \ref{sec:method}) with the observed emission line
fluxes. We mainly study the results for the \ha\ line because we have \ha\ measurements for 137 galaxies allowing for significant statistics. Furthermore, the \niilam\ to \hb\ flux ratio depends strongly on metallicity with $F_{\niilam}/F_{\hb}=0.175$ for $Z=0.004$ and $F_{\niilam}/F_{\hb}=0.404$ for $Z=0.02$, while the \oiiilam\ to \hb\ flux ratio is less metallicity dependent \citep[$F_{\oiiilam}/F_{\hb}=4.752$ and $F_{\oiiilam}/F_{\hb}=4.081$ for $Z=0.004$ and $Z=0.02$, respectively,][]{AF03}, but more sensitive to the \hii\ physical conditions like ionization parameter or leakage of ionizing photons \citep[e.g.,][]{kewleyetal2013a,nakajimaouchi14}.

     \begin{figure}[htbf]
     \centering
     \includegraphics[width=8cm,trim=0.5cm 0cm 1.7cm 1.25cm,clip=true]{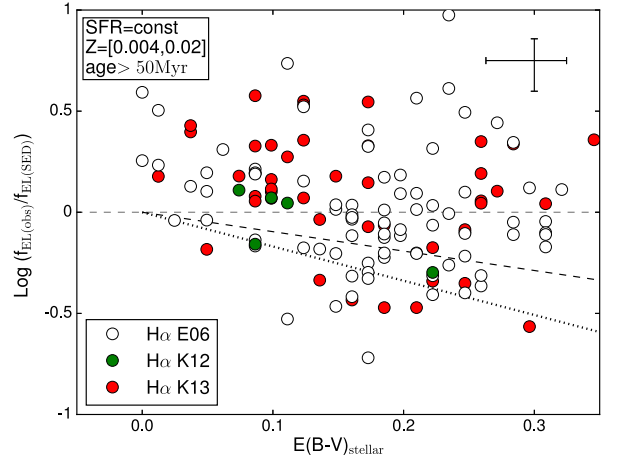}
     \caption{\dfel\ vs. E(B-V)$_{\mathrm{stellar}}$ for the entire
       sample.  The dashed
       line and the dotted line show the expected relation from
       equation \ref{eq:cardha} and equation \ref{eq:calha},
       respectively.
       }
     \label{fig:Del_ebv}
     \end{figure}
     
     \begin{figure}[htbf]
     \centering
     \includegraphics[width=8cm,trim=0.5cm 0cm 1.7cm 1.25cm,clip=true]{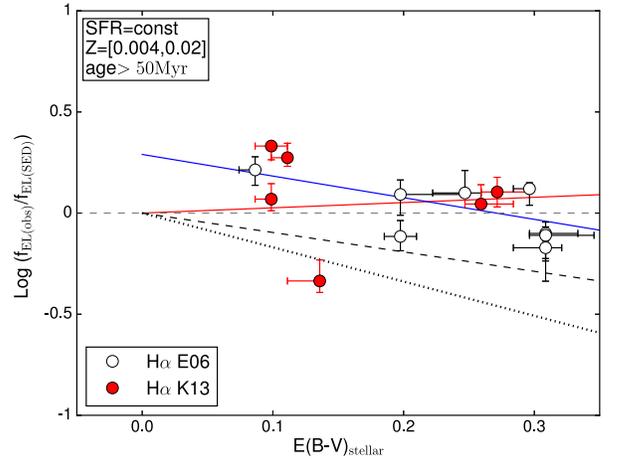}
     \caption{Same as Fig.~\ref{fig:Del_ebv} for the MIPS detected sample. We show 2 different fits to the \ha\ data: in blue, we show a linear fit to the data, with the slope and intercept as free parameters. Additionally, in red, we show the linear fit where we fix the intercept to be zero. 
       }
     \label{fig:Del_ebv_mips}
     \end{figure}
 
 To quantify this comparison,
we computed $\Delta\mathrm{f}_\mathrm{EL}=\log(\mathrm{f}_\mathrm{EL(obs)}/\mathrm{f}_\mathrm{EL(SED)})$,
where $\mathrm{f}_\mathrm{EL(obs)}$ is the observed emission line flux and $\mathrm{f}_\mathrm{EL(SED)}$ is the SED predicted emission line flux. We remind the reader that the SED fitting assumes that the color excesses of the nebular regions and the
stellar continuum are equal, i.e., $\mathrm{E(B-V})_{\mathrm{stellar}} = \mathrm{E(B-V})_{\mathrm{nebular}}$ (derived assuming the Calzetti extinction curve). Under this assumption, the \ha\ extinction is
$\mathrm{A}_{\ha}=3.33 \times\mathrm{E(B-V)}_{\mathrm{stellar}}$.
The comparison between our predicted and observed fluxes for the 
entire sample is shown in Figure \ref{fig:Del_mips}. We assume a constant SFH because this SFH provides the best match to the $z\sim2$ galaxy physical properties (Sect.~\ref{sec:genprop}). On average, our method used to model nebular emission from the photometry provides \ha\ fluxes consistent with observations: the median shows a slight underprediction of the flux by $\sim3\%$ with $\sigma=0.31$ dex.

We first study
how
$\Delta\log\mathrm{f_{\mathrm{EL}}}$
correlates with the stellar color excess (Figure
\ref{fig:Del_ebv}). We compare our results with the relationship between nebular and color excess for local galaxies\footnote{Equation \ref{eq:extraatt} is often misinterpreted as both color
excesses being derived with the same attenuation curve, but the intent was that the nebular regions are described by a line-of-sight
(e.g., Milky Way) extinction curve \citep[e.g.,][]{steideletal2014,reddyetal2015}.} from \cite{calzettietal2000}:
 \begin{equation}
\mathrm{E(B-V})_{\mathrm{stellar}} = (0.44\pm0.03)\times\mathrm{E(B-V})_{\mathrm{nebular}}
\label{eq:extraatt}
\end{equation}

The extinction at a wavelength $\lambda$
is related to the color excess $\mathrm{E(B-V)}$ and to reddening curve k($\lambda$) by $A_\lambda=k(\lambda)\times\mathrm{E(B-V)}$. At the \ha\ wavelength (6562.8\AA), we have $k_\mathrm{Calzetti}(\ha)=3.33$ and $k_\mathrm{Cardelli}(\ha)=2.52$, which leads respectively in terms of \ha\ extinction (assuming Equation \ref{eq:extraatt}):
 \begin{equation}
\mathrm{A}_{\ha,\mathrm{Cardelli}}=5.72 \times \mathrm{E(B-V)}_{\mathrm{stellar}}
\label{eq:cardha}
\end{equation}
 \begin{equation}
\mathrm{A}_{\ha,\mathrm{Calzetti}}=7.55 \times \mathrm{E(B-V)}_{\mathrm{stellar}}
\label{eq:calha}
\end{equation}

\begin{figure}[htbf]
\centering
\includegraphics[width=8cm,trim=0.5cm 0cm 1.5cm 1.25cm,clip=true]{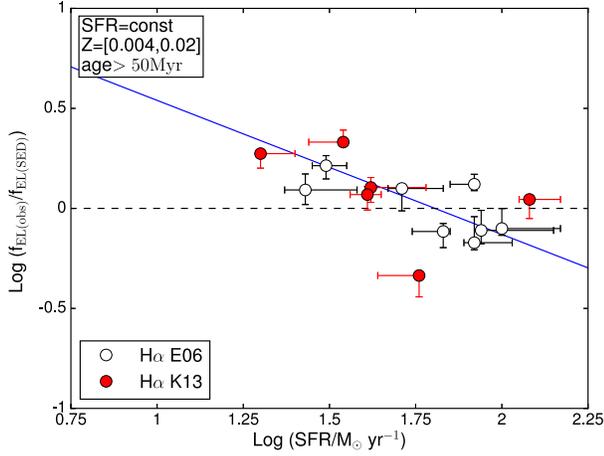}
\caption{\dfel\ vs. SFR for the MIPS detected sample. Linear fit to the data is shown with the blue line.}
\label{fig:Del_sfr}
\end{figure} 

Our comparison between
predicted fluxes (based on \sfrsed\ derived assuming energy balance) and observed fluxes should allow us to discriminate between equations \ref{eq:cardha} and \ref{eq:calha}.
If the relative color excess of the UV continuum and the nebular emission lines at
$z\sim2$ is similar to the one derived in local starburst galaxies, we expect to
overpredict emission line fluxes because we assume
$\mathrm{A}_\mathrm{V,nebular}=\mathrm{A}_\mathrm{V,stellar}$. Furthermore, if the nebular color excess is related to UV color excess as in local starburst galaxies (i.e., $\mathrm{E(B-V)}_\mathrm{nebular}=\mathrm{constant}\times\mathrm{E(B-V)}_\mathrm{stellar}$), we should find an  anticorrelation between \dfel\ and $\mathrm{E(B-V)}_\mathrm{stellar}$.
If we assume, as we do in
our SED fitting procedure, that the same attenuation curve applies to
both nebular and stellar emission, we get:
 \begin{equation}
\frac{\mathrm{f}_\mathrm{EL(obs)}}{\mathrm{f}_\mathrm{EL(SED)}}=\frac{\mathrm{f}_\mathrm{intrinsic}\times10^{-0.4\times A_{\lambda\mathrm{(obs)}}}}{\mathrm{f}_\mathrm{intrinsic}\times10^{-0.4\times A_{\lambda\mathrm{(SED)}}}}
\end{equation}
 \begin{equation}
\log\left(\frac{\mathrm{f}_\mathrm{EL(obs)}}{\mathrm{f}_\mathrm{EL(SED)}}\right)=-0.4\times (A_{\lambda(\mathrm{obs})}-A_{\lambda(\mathrm{SED})})
\end{equation}
 \begin{equation}
\Delta\log\mathrm{f}_\mathrm{EL}=-0.4\times k(\lambda) \times(\mathrm{E(B-V)}_\mathrm{neb}-\mathrm{E(B-V)}_\mathrm{stel})
\label{eq:corel}
\end{equation}
A Spearman correlation test performed on data from Figure~\ref{fig:Del_ebv}
 leads to $\rho=-0.31$ with standard deviation from null hypothesis $\sigma=3.63$ (Table \ref{tab:spear}), which indicates that \dfel\ and E(B-V)$_\mathrm{stellar}$ are possibly anticorrelated.
We also compare the results expected from equation
\ref{eq:cardha} and equation \ref{eq:calha} with our data. While our method is able to reproduce observed emission lines within a factor $\lesssim2$, uncertainties remain too large to derive a possible relation between E(B-V)$_\mathrm{stellar}$ and \dfel. To decrease the uncertainties, we focus on the MIPS detected sample (Fig.~\ref{fig:Del_ebv_mips}). We compare the results expected from equation
\ref{eq:cardha} and equation \ref{eq:calha} with linear fits to our
data:
 \begin{equation}
\dfel=a\times\mathrm{E(B-V)}_\mathrm{stellar}+b
\label{eq:linfit}
\end{equation}
and we perform two linear fits:
\begin{itemize}
\item (1) We fit the MIPS detected sample (blue in Figure \ref{fig:Del_ebv}).
\item (2) We fit the MIPS detected sample assuming $b=0$ in equation \ref{eq:linfit} (red).
\end{itemize}
\begin{figure}[htbf]
\centering
\includegraphics[width=8cm,trim=0.5cm 0cm 1.5cm 1.25cm,clip=true]{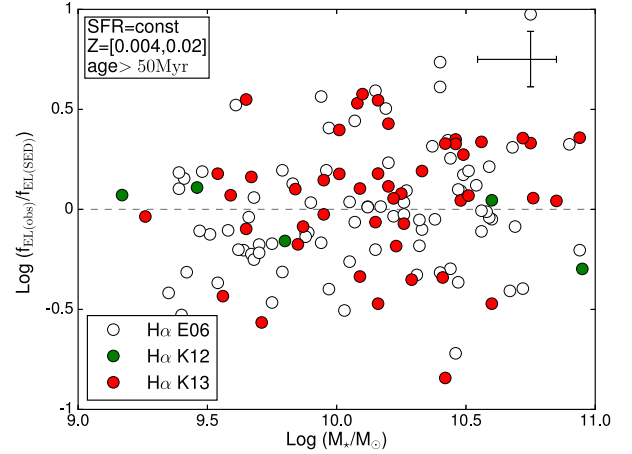}
\caption{\dfel\ vs. \mstar\  for the entire sample. Typical error bar is shown in the top right corner.}
\label{fig:Del_mstar}
\end{figure} 
Assuming $b=0$ means that we consider that our SED fitting code is able to reproduce the emission line fluxes for dust free galaxies.  The intercept of fit (1) means that we underestimate observed emission line fluxes for dust free galaxies by a factor $\sim2$. When assuming $b=0$,$\mathrm{E(B-V)}_\mathrm{stellar}$ and $\mathrm{E(B-V)}_\mathrm{nebular}$ are similar (assuming a Calzetti extinction curve for both emission). This result is consistent with the conclusion from \cite{shivaeietal2015a}. However, fit (1) shows that we are not predicting exactly emission fluxes for galaxies with no extinction. If we assume that there is a systematic offset between predicted and observed fluxes, not dependent on E(B-V)$_\mathrm{stellar}$, we can compare the slope of our relation with expectation from equation \ref{eq:cardha} and \ref{eq:calha}. Then, our slope is consistent with equation \ref{eq:cardha}.


 \subsection{Relation between nebular color excess and other physical properties}
 \label{sec:corr}
 
In this section we determine wether \dfel\ possibly correlates with other physical properties such as SFR, \mstar, and sSFR. We summarize the correlation test and their associated significances for the different samples in Table~\ref{tab:spear}.

We find an anticorrelation between log SFR and \dfel\ with a level of confidence $>6\sigma$
($\rho=-0.53$, $\sigma=6.67$).
This relation could explain the discrepancies among different previous studies. We derive  the relation between \dfel\ and log SFR using the MIPS detected sample because uncertainties are too large when IR information is missing.
Assuming a constant SFH, the relation from Figure \ref{fig:Del_sfr} implies:
 \begin{eqnarray}
\mathrm{E(B-V)}_\mathrm{stellar}-\mathrm{E(B-V)}_\mathrm{nebular}=\nonumber\\(-0.50\pm0.10)\times\log \mathrm{SFR}+(0.91\pm0.18)
\label{eq:extrasfr}
\end{eqnarray}
This anticorrelation between \dfel\ and log SFR implies that for $z\sim2$ star-forming galaxies with $\log\mathrm{SFR}\le1.82^{+0.90}_{-0.60}$, we have $\mathrm{E(B-V)}_\mathrm{stellar}\sim\mathrm{E(B-V)}_\mathrm{nebular}$. 
While uncertainty is large, this result is consistent with the SFR value derived in \cite{reddyetal2015}. We discuss this point further in Section \ref{sec:other}. 

\begin{figure}[htbf]
\centering
     \includegraphics[width=8cm,trim=0.5cm 0cm 1.7cm 1.25cm,clip=true]{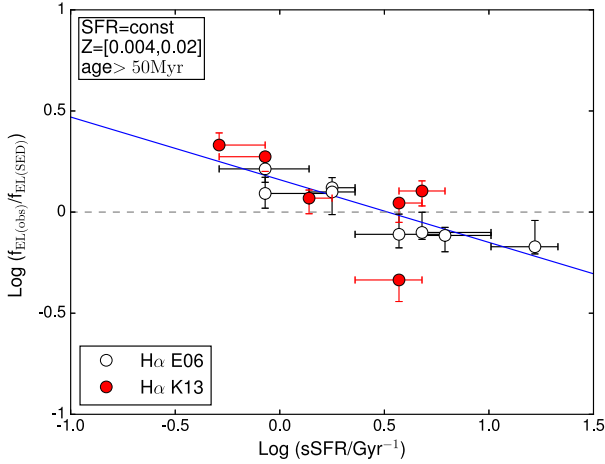}
     \caption{\dfel\ vs. sSFR for the MIPS detected sample. Linear fit to the data is shown with the blue line.}
     \label{fig:Del_ssfr}
     \end{figure}

We find no correlation betwen \mstar\ and \dfel\ (Fig.~\ref{fig:Del_mstar}), with $\rho=0.16$ and $\sigma=1.86$. On the contrary, we find an anticorrelation between sSFR and \dfel\ with $\rho=-0.52$ and $\sigma=6.48$ (Fig.~\ref{fig:Del_ssfr}). Again, the uncertainties are too large to derive a relation between \dfel\ and log sSFR for the entire sample, and we use the MIPS detected sample to derive a relation. From Figure~\ref{fig:Del_ssfr}, we get:
 \begin{eqnarray}
\mathrm{E(B-V)}_\mathrm{stellar}-\mathrm{E(B-V)}_\mathrm{nebular}=\nonumber\\(-0.31\pm0.07)\times\log \mathrm{(sSFR)}+(0.16\pm0.03)
\label{eq:extrassfr}
\end{eqnarray}

We conclude that the analysis of our sample supports marginally the relation derived between $\mathrm{E(B-V)}_\mathrm{stellar}$ and $\mathrm{E(B-V)}_\mathrm{nebular}$ in local star-forming galaxies \citep[Eq.~\ref{eq:cardha},][]{calzettietal2000}. While we do not find a correlation between the stellar mass and $\mathrm{E(B-V)}_\mathrm{stellar}-\mathrm{E(B-V)}_\mathrm{nebular}$, we find anticorrelations between SFR/sSFR and $\mathrm{E(B-V)}_\mathrm{stellar}-\mathrm{E(B-V)}_\mathrm{nebular}$ \citep{reddyetal2015}.

\begin{table*}[htbf]
         \centering
         \caption{Spearman correlation test between \dfel\ and physical parameters}
         \begin{tabular}{lcccccc}
           \hline
           \hline
           Physical parameters & \multicolumn{2}{c}{MIPS detected ($N=14$)} & \multicolumn{2}{c}{MIPS sample ($N=34$)} & \multicolumn{2}{c}{Entire sample ($N=137$)}\\
           & $\rho^{\mathrm{a}}$ & $\sigma^{\mathrm{b}}$ & $\rho$ & $\sigma$ & $\rho$ & $\sigma$\\
           \hline
           E(B-V)$_\mathrm{stellar}$  &  -0.46  &  1.61  &  -0.35  &  1.96  & -0.31 & 3.63 \\
           SFR & -0.62 & 2.31 & -0.78 & 5.68 & -0.53 & 6.67 \\
            \mstar & 0.66 & 2.58 & 0.13 & 0.73 & 0.16 & 1.86 \\
            sSFR & -0.76 & 3.21 & -0.57 & 3.49 & -0.52 & 6.48\\
           \hline
         \end{tabular}
         \label{tab:spear}
         \begin{list}{}{}
\item[$^{\mathrm{a}}$ Spearman rank correlation coefficient.]
\item[$^{\mathrm{b}}$ Standard deviations from null hypothesis.]
\end{list}
      \end{table*}


\section{Discussion}
\label{sec:discuss}

\subsection{Modeling of nebular emission}
\label{sec:modneb}

One result of our study is that our SED fitting code is able to reproduce consistently the broad-band photometry and emission line fluxes, within their respective uncertainties, if we assume energy balance. This result is not a trivial finding as our modeling of the nebular emission relies on several assumptions regarding the Lyman continuum escape fraction, ISM physical conditions, empirical ratios between emission lines, and the IMF.

\begin{figure}[htbf]
\centering
     \includegraphics[width=8cm,trim=0.5cm 0cm 1.5cm 1.25cm,clip=true]{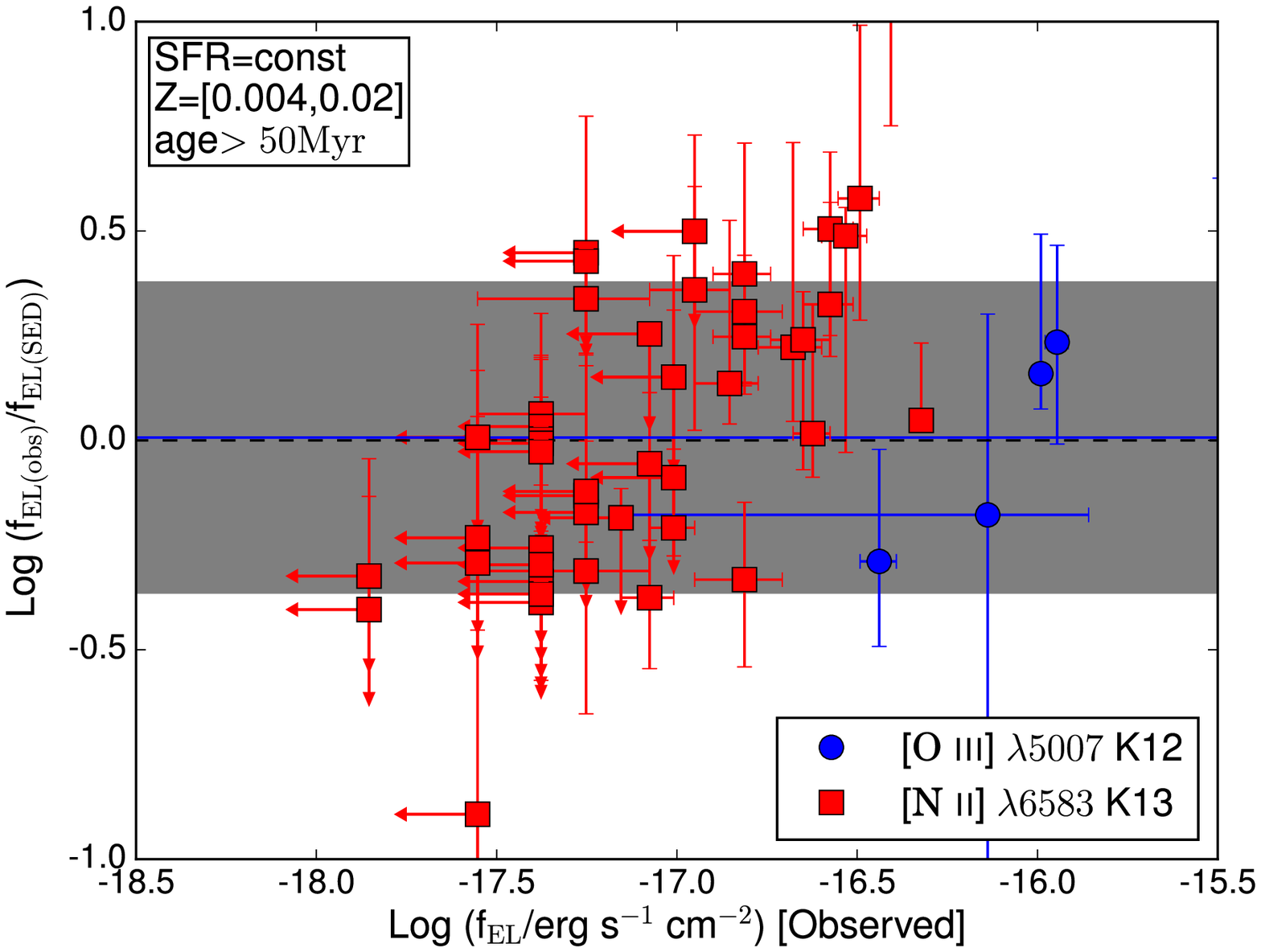}
     \caption{Same as Figure~\ref{fig:Del_mips} for the \niilam\ and \oiiilam\ measurements. The blue line shows the median \dfel\ value and the grey area the 1$\sigma$ dispersion.}
     \label{fig:Del_other}
     \end{figure}

However, we can not rule out that a different set of physical properties could produce similar emission line fluxes. For example, the \ha\ flux would decrease by a factor 1.8 if we consider a Chabrier IMF  \citep{chabrier2003} instead of a Salpeter IMF.
Furthermore, the absolute flux of hydrogen emission lines are directly proportional  to the number of ionizing photons per second, and so to the Lyman continuum escape fraction \citep[e.g.,][]{krueger+95}. Nevertheless, we show in Figure~\ref{fig:Del_other} the comparison between predicted and observed \oiiilam\ and \niilam\ fluxes. We have shown in the previous sections that the effect of difference between nebular and stellar color excesses is negligible in the range of SFR (and sSFR) probed with our sample. Therefore, \dfel\ for these two lines is mostly sensitive to the metallicity for \niilam\ \citep[from \zsun\ to 0.2\zsun\, the \niilam/\ha\ ratio decreases by a factor 2.34,][]{AF03}, and to the ISM physical conditions for \oiiilam \citep[e.g.,][]{kewleyetal2013a}. Our method is able to reproduce these lines with a similar accuracy as it reproduces \ha, with a median \dfel$\sim0$ and $\sigma=0.34$ dex.

        \begin{figure}[htbf]
\centering
     \includegraphics[width=8cm,trim=0.5cm 0cm 1.7cm 1.25cm,clip=true]{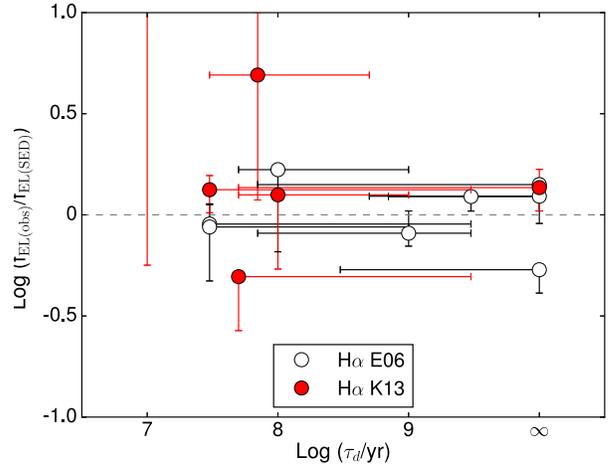}
     \caption{\dfel\ vs. timescale $\tau_d$ for the MIPS detected sample assuming a declining star-formation history. We show the result only for the \ha\ data for more clarity.}
     \label{fig:Del_tau}
     \end{figure} 
     
We show that the method used in this work to model nebular emission \citep{schaerer&debarros2009,schaerer&debarros2010} is able to reproduce \ha, \niilam, and \oiiilam. This is an important result because of the implications for higher redshift studies ($z>3$), where emission lines can not be measured with current instruments while observed SEDs exhibit evidences of impact of nebular emission \citep[e.g.,][]{starketal2013,debarrosetal2014,nayyeri+14,marmol+15,smitetal2015}.

  
 \subsection{Star-formation history}
 \label{sec:sfh}
 
Several studies have used the comparison between direct star-formation tracers and SFR(SED) to put constraints on star-formation timescales. Their main conclusion is that declining SFHs result in SFRs(SED) that are inconsistent with those inferred from multi-wavelength tracers of star formation (e.g., the sum of the IR and UV), or that it is necessary to assume a lower limit on the decay timescale \citep[$\tau_d\gtrsim300\mathrm{Myr}$;][]{wuytsetal2011,reddyetal2012a,priceetal2014}. To gain insight on the star-formation timescales for both rising and declining SFHs, we rely on the MIPS detected sample since the SFR uncertainties are much lower when L(IR) is used to perform SED fitting assuming energy balance (the uncertainties derived from SED fitting are 0.1 dex versus 0.35 dex without assuming energy balance). Therefore, we use both IR and emission lines to place constraints on the SFH timescale \citep{schaereretal2013}.

 \begin{figure}[htbf]
\centering
     \includegraphics[width=8cm,trim=0.5cm 0cm 1.7cm 1.25cm,clip=true]{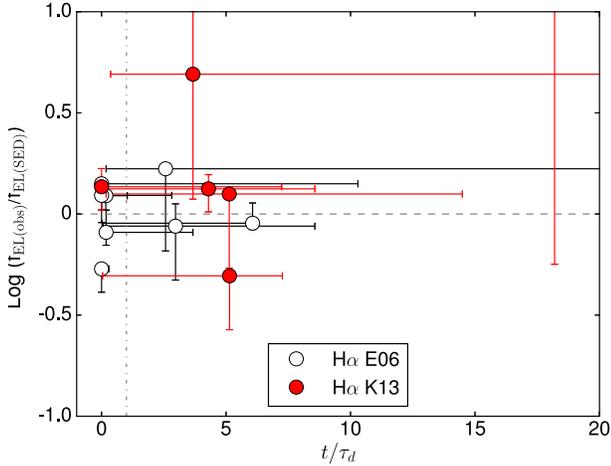}
     \caption{\dfel\ vs. the ratio between age and timescale $t/\tau_d$ for the MIPS detected sample assuming a declining star-formation history and and a relation between E(B=V)$_\mathrm{nebular}$ and E(B=V)$_\mathrm{stellar}$ described by Equation~\ref{eq:extrasfr}. The dash-dotted line shows $t/\tau_d=1$. We show the result only for the \ha\ data for more clarity}
     \label{fig:Del_ttau}
     \end{figure} 

Since we assume energy balance, the IR luminosity predicted from SED fitting are consistent with L(IR)$_{24\mu\mathrm{m}}$ by construction. The difference between predicted and observed emission lines fluxes assuming a declining SFH do not differ significantly from other SFHs, except for galaxies with blue stellar color excess. In Figure~\ref{fig:Del_tau}, we show the relation between \dfel\ and the timescale $\tau_d$ and in Figure~\ref{fig:Del_ttau}, the relation between \dfel\ and the ratio between age and timescale $t/\tau_d$. Overall, the SED fitting reproduces the observed \ha\ fluxes with the same accuracy as for the entire sample (i.e., the fluxes are reproduced within a factor 2). There are two galaxies for which the fluxes are significantly underpredicted (\dfel>0.5) and they are fit with short timescales ($\leq100$Myr).
In Figure~\ref{fig:Del_ttau}, we see that the age to timescale ratio can be as high as $\sim5$ with the predicted emission line fluxes still consistent with the observed values.
The two galaxies for which we significantly underpredict the emission line fluxes have the lowest IR luminosities of the sample ($\leq10^{11}\lsun$). If we add the \ha\ flux measurement constraints to the SED fit for these two galaxies (excluding solutions providing \ha\ fluxes inconsistent with the observed ones within $1\sigma$), timescales are $\tau_d\geq500\mathrm{Myr}$ and the age to timescale ratios are $t/\tau_d<1$. Our conclusions regarding declining SFHs are consistent with the literature \citep[e.g.,][]{wuytsetal2011,priceetal2014} but for a fraction of our MIPS detected sample ($\sim1/3$), we find that galaxy SEDs are fit with short timescales ($30\mathrm{Myr}\leq\tau_d\leq1\mathrm{Gyr}$) and relatively large $t/\tau_d$ ratios ($0\leq t/\tau_d\leq5$), and have predicted physical quantities (L(IR) and \ha\ flux) consistent with observations.

We overpredict the emission line fluxes for objects fit with a short exponential timescale ($\tau_r<10^8\mathrm{yr}$) for the rising SFH (data not shown), while we underpredict fluxes for fit with a large timescale ($>10^9\mathrm{yr}$).

   \begin{figure}[htbf]
     \centering
     \includegraphics[width=8cm,trim=0.25cm 0.1cm 1.6cm 1.2cm,clip=true]{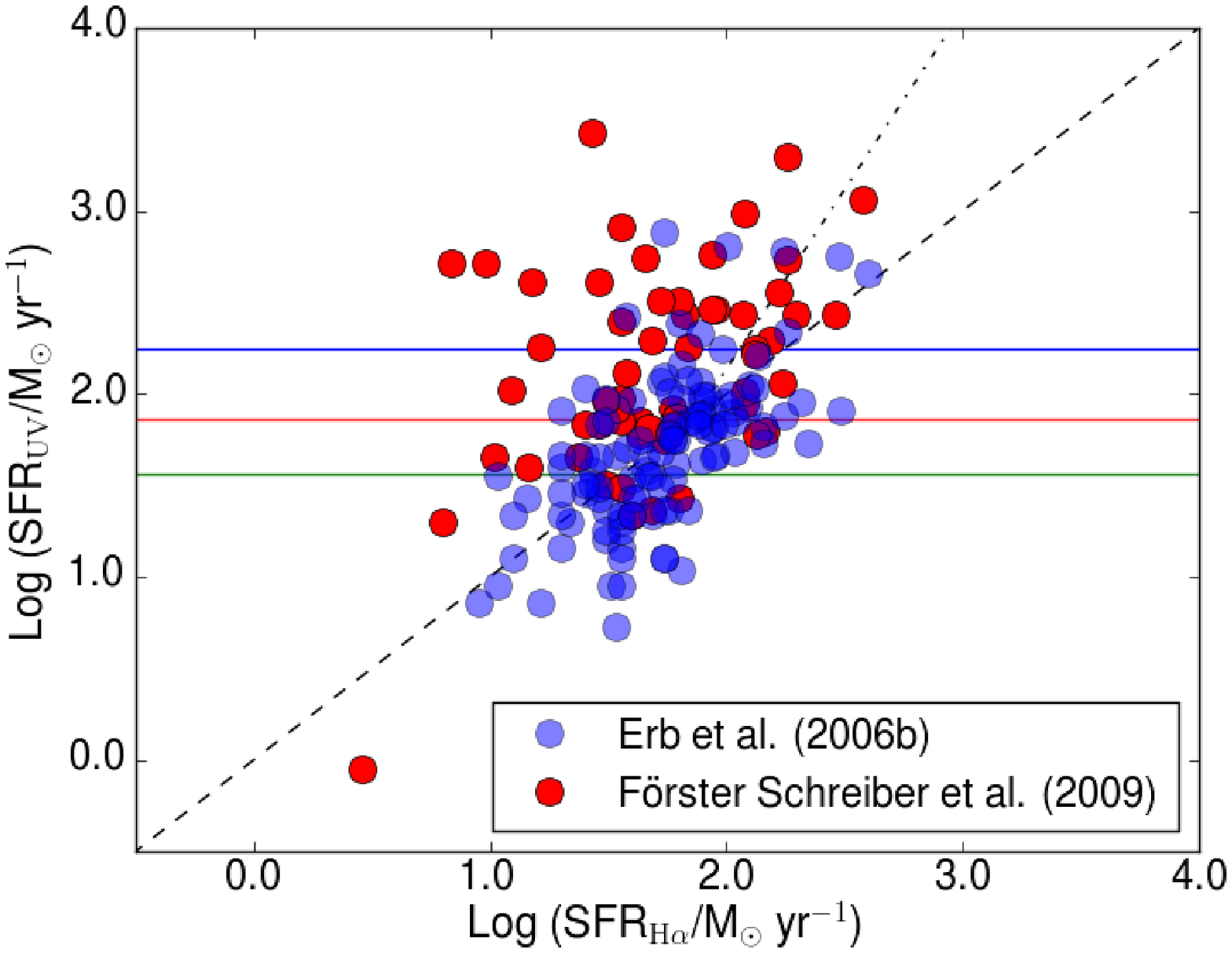}
     \caption{SFR$_{\ha}$ vs. SFR$_{\mathrm{UV}}$ assuming $\mathrm{E(B-V)}_\mathrm{stellar}=\mathrm{E(B-V)}_\mathrm{nebular}$ for the \cite{erbetal2006b} and \cite{FS09} samples. The horizontal red line shows $\log\hspace{1mm}\mathrm{SFR}=1.82$ (Section~\ref{sec:corr}, Equation~\ref{eq:extrasfr}), the blue line $\log\hspace{1mm}\mathrm{SFR}=2.25$ \citep{yoshikawaetal2010}, and the green line $\log\hspace{1mm}\mathrm{SFR}=1.56$ \citep[][converting the value from a Chabrier to a Salpeter IMF]{reddyetal2015}.
The dash-dotted line is the expected log SFR(\ha)-log SFR(UV) relation assuming that $\mathrm{E(B-V)}_\mathrm{nebular}-\mathrm{E(B-V)}_\mathrm{stellar}$ is related to log SFR as described in Equation~\ref{eq:extrasfr}.
The dashed line is the one to one relation.}
     \label{fig:comp_extra}
 \end{figure} 

To perform SED fitting and get results consistent with observed emission line fluxes, assuming energy balance (and so having IR information), we show that the timescale used with a declining SFH is $\tau_d\geq30\mathrm{Myr}$ and that the age to timescale ratio is $\leq5$. For a rising SFH, the timescale is in the range $300\mathrm{Myr}\leq\tau_r\leq1\mathrm{Gyr}$.


\subsection{Comparison with other studies}
\label{sec:other}

We find an anticorrelation between SFR and \dfel\, which we interpret as a correlation between SFR and $\mathrm{E(B-V})_{\mathrm{stellar}}-\mathrm{E(B-V})_{\mathrm{nebular}}$ (Equation~\ref{eq:extrasfr}): the nebular color excess becomes larger than the stellar color excess with increasing SFR.
As pointed out by \cite{reddyetal2010}, previous studies included galaxies with different ranges in SFR and \mstar. Therefore we check now if our relation found between SFR and \dfel\ could explain the discrepant results on this matter.

We compare the \cite{erbetal2006b} sample, for which the nebular and stellar color excesses are similar, and the \cite{FS09} sample, for which the nebular color excess is higher than the stellar color excess (Equation~\ref{eq:extraatt}). It is worth noting that a fraction of our sample is drawn from the \cite{erbetal2006b} sample.

We show in Figure \ref{fig:comp_extra} the comparison between SFR$_\mathrm{UV}$ and SFR$_{\ha}$ for these two studies assuming $\mathrm{E(B-V)}_\mathrm{stellar}=\mathrm{E(B-V)}_\mathrm{nebular}$, a Calzetti attenuation curve, and after converting the SFRs from a Chabrier IMF to a Salpeter IMF. We expect to see an increasing  discrepancy between SFR$_{\ha}$ and SFR$_\mathrm{UV}$ above the value derived from Equation \ref{eq:extrasfr} with $\log\mathrm{SFR}=1.82$. We also show similar SFR thresholds found in \cite{yoshikawaetal2010} and \citep{reddyetal2015}.

Because we do not have Balmer decrement measurements in order to correct \ha, we consider the dust-corrected SFR$_\mathrm{UV}$ as the one with which we compare SFR$_\mathrm{\ha}$. We can predict the difference between SFR(\ha) derived assuming $\mathrm{E(B-V)}_\mathrm{stellar}=\mathrm{E(B-V)}_\mathrm{nebular}$ and SFR(\ha) derived assuming the relation between SFR and the difference in the color excesses as in Equation \ref{eq:extrasfr} (dash-dotted line in Figure~\ref{fig:comp_extra}). At log SFR$=1.82$, this difference is zero and reach 0.56 dex at $\mathrm{SFR(UV)}_\mathrm{corrected}=10^3\msunyr$. This explains why we do not see any significant discrepancy between SFR(UV) and SFR(\ha) in Figure~\ref{fig:comp_extra} below the SFR limit derived in \cite{yoshikawaetal2010}. Assuming $\mathrm{E(B-V)}_\mathrm{stellar}=\mathrm{E(B-V)}_\mathrm{nebular}$ leads to underpredict $\mathrm{SFR}_{\ha}$ in comparison with $\mathrm{SFR}_\mathrm{UV}$ above the threshold found in \cite{yoshikawaetal2010}. Therefore we conclude that the discrepant result about the difference between nebular and stellar color excesses  is likely a selection effect, the \cite{FS09} sample probing higher SFRs. The apparent discrepancy between Equation~\ref{eq:extrasfr} and the \cite{FS09} sample can be explained by the large dispersion in color excess for individual galaxies \citep{reddyetal2015}.


\section{Conclusions}
\label{sec:conclusion}    

We use a sample of 149 star-forming galaxies spectroscopically confirmed at $z\sim2$ with optical emission line measurements to investigate the relative dust attenuation of the UV continua and the nebular emission in these galaxies. For a subsample of 41 galaxies, MIPS 24\micron\ data are available and we use them to derive the infrared luminosity of these galaxies. We compare all the available observed/derived properties of our sample (emission line flux, IR luminosity) with predictions from stellar population synthesis models, taking into account nebular emission, and assuming energy balance between UV and IR emission. We explore a large parameter space, exploring different star-formation histories (constant, declining, rising) and the impact of different attenuation curves (Calzetti and SMC curves) on our results.

We compare the properties (stellar masses, SFRs) of our sample with properties found in literature \citep{daddietal2007,reddyetal2012b} to ensure the representativeness of our sample. It appears that we are not able to reproduce the observed UV $\beta$ slope for the ULIRGs in our sample when we assume energy balance and we exclude these 7 galaxies from our analysis. Our results confirm that the Calzetti curve is not appropriate for ULIRGs (Appendix~\ref{sec:ulirg}) and an appropriate attenuation curve should be define to derive properly their physical properties \citep{reddyetal2010,caseyetal2014}.

Our conclusions are as follow:
 \begin{itemize}
 	\item Assuming energy balance, our SED fitting code is able to reproduce the observed emission line fluxes for galaxies in our sample (within a factor $<2$).
		\item Young galaxies ($\leq100\hspace{1mm}\mathrm{Myr}$) are statistically better fit with an SMC curve rather than a Calzetti curve \citep{reddyetal2010}. 
	\item We find a correlation between SFR and $\mathrm{E(B-V)}_\mathrm{stellar}-\mathrm{E(B-V)}_\mathrm{nebular}$, and also between sSFR and $\mathrm{E(B-V)}_\mathrm{stellar}-\mathrm{E(B-V)}_\mathrm{nebular}$. The color excess difference increases with increasing SFR and sSFR. The correlation between the amount of extra attenuation toward nebular emission and the SFR could explain discrepancies among previous studies about the nebular attenuation relative to stellar attenuation \citep[e.g.,][]{erbetal2006b,FS09,reddyetal2010}.
	\item We use SED, IR and emission line flux measurements to constrain the star formation history. Overall, a large range of SFHs can reproduce the observables. We find that assuming a declining SFH, the timescale is $\tau_d>30\mathrm{Myr}$ and the age to timescale ratio is $t/\tau_d\leq5$. For a rising SFH, the timescale is $300\mathrm{Myr}\leq\tau_r\leq1\mathrm{Gyr}$.
\end{itemize}

We look forward to comparing our results with current ongoing survey \citep{reddyetal2015}. This will also allow to determine much more precisely the ratio $\mathrm{E(B-V)}_\mathrm{stellar}/\mathrm{E(B-V)}_\mathrm{nebular}$, and determine if there is an evolution of this ratio from low to high redshift. However, our results suggest that this kind of observations must aim a large range of SFRs and stellar masses to avoid any bias. It would also be interesting to measure Balmer decrement of ULIRGs to derive the best method to infer their physical properties such as stellar mass, SFR, and dust attenuation.


\acknowledgments

We thank the referee for suggestions that clarify the text and the analysis. S. D. B. was supported by a fellowship from the Swiss National Science Foundation. We thank the referee for suggestions that clarify the text and the analysis. We acknowledge Chuck Steidel and his group for help in obtaining the data presented here. We thank Alice Shapley, Valentino Gonz\'alez, Pascal Oesch, Bahram Mobasher, Daniel Schaerer, Panos Sklias, Alberto Dom\'inguez, and Matthew Hayes for useful discussions about this work.  We also thank Roser Pell\'o, Daniel Schaerer, and Panos Sklias for their support with {\it HyperZ}.


\begin{appendices}

\section{ULIRG physical properties}
\label{sec:ulirg}

   \begin{figure*}[htbf]
    \centering
     \includegraphics[width=8.75cm,trim=0cm 0cm 0cm 0cm,clip=true]{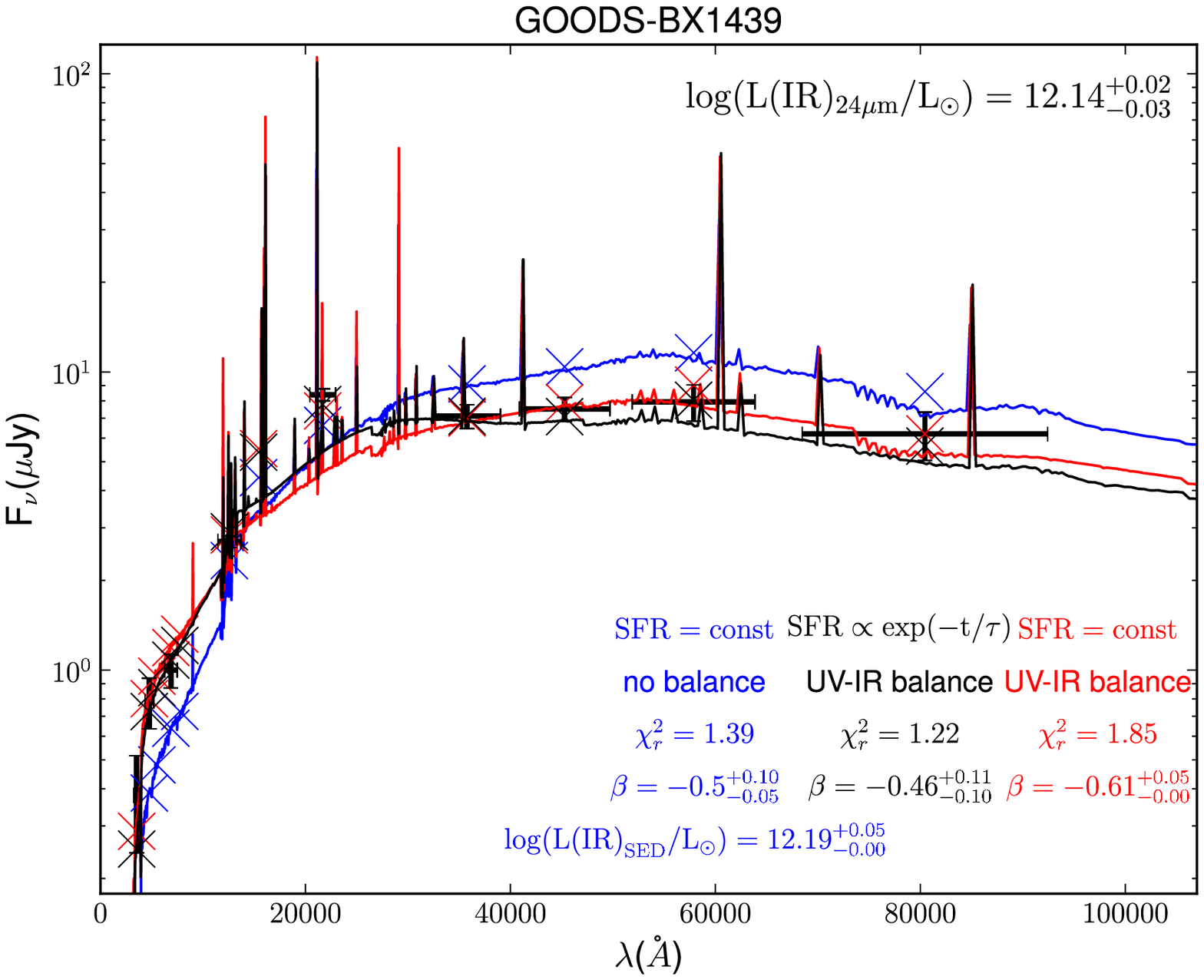}
     \includegraphics[width=8.75cm,trim=0cm 0cm 0cm 0cm,clip=true]{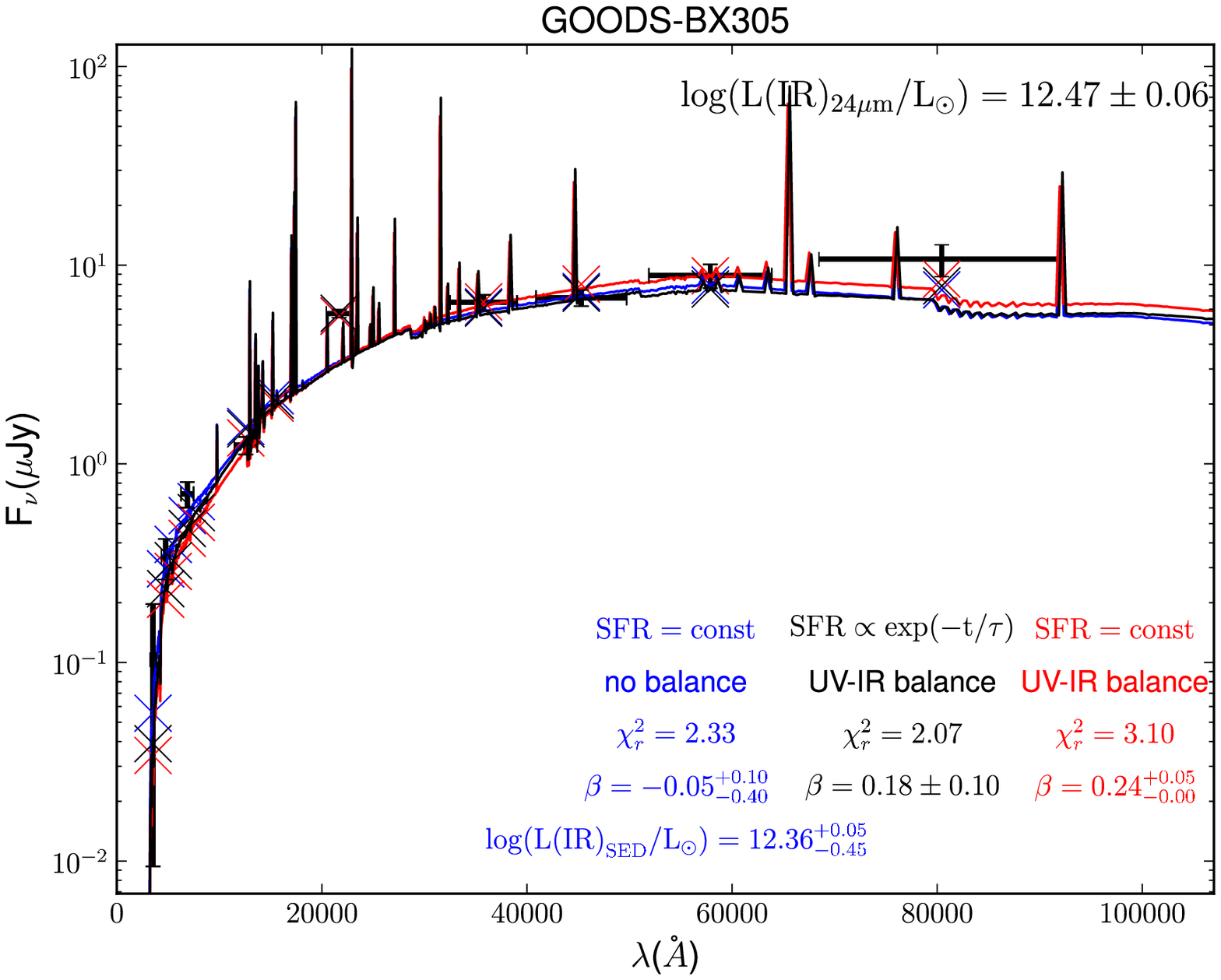}
     \includegraphics[width=8.75cm,trim=0cm 0cm 0cm 0cm,clip=true]{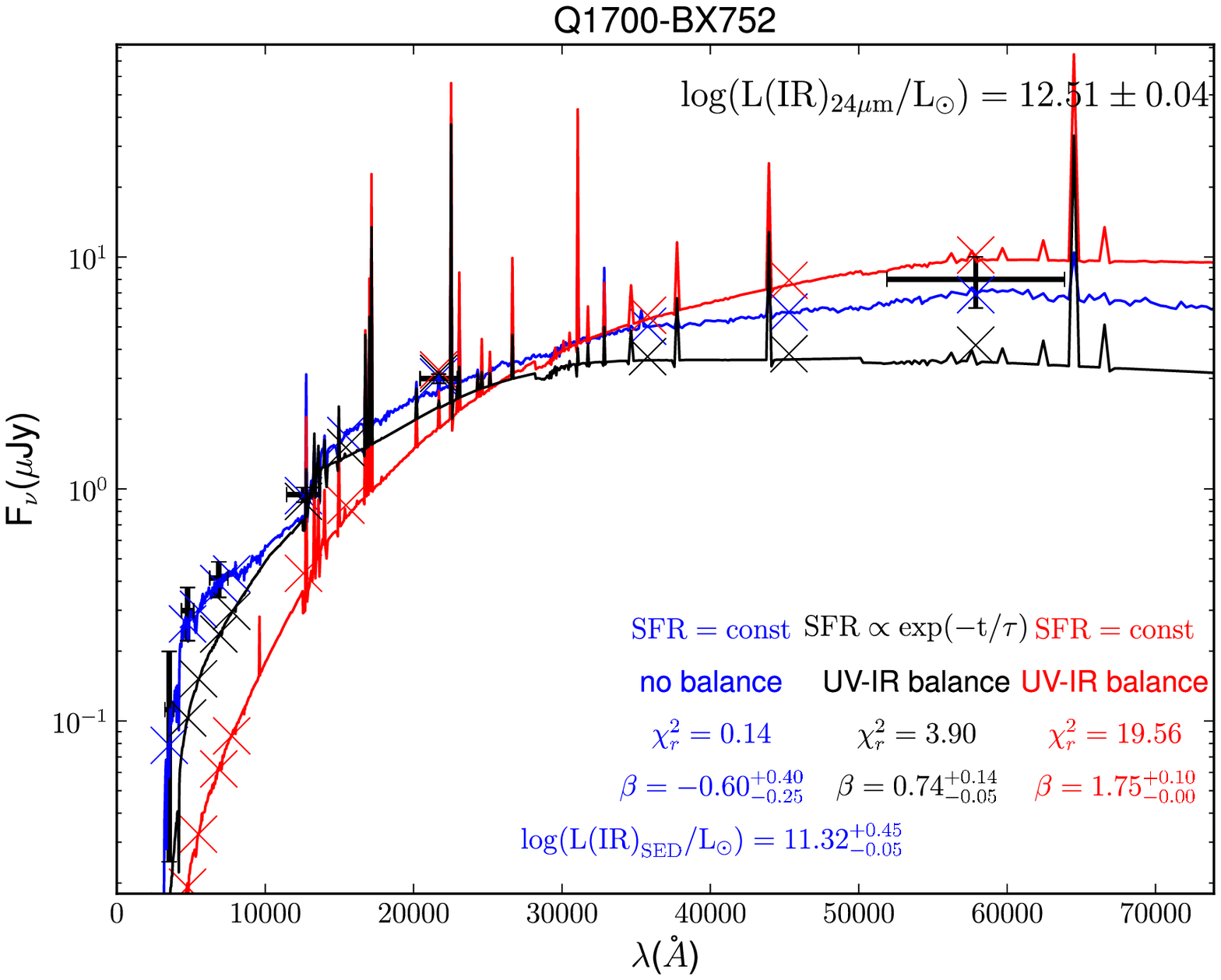}
     \includegraphics[width=8.75cm,trim=0cm 0cm 0cm 0cm,clip=true]{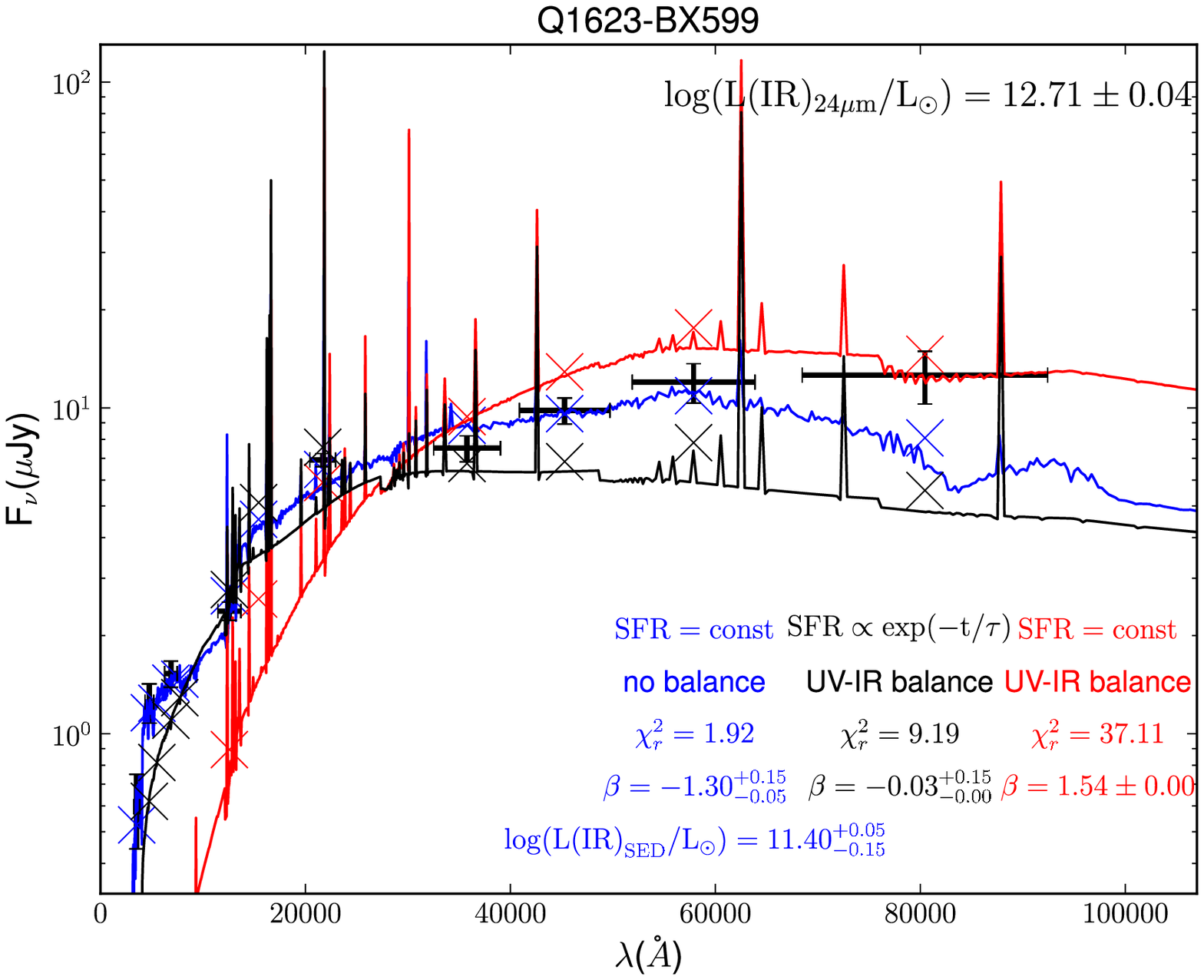}
          \caption{Observed SEDs (black thick points) of the seven ULIRGs in our sample with the best-fits for three models: constant SFH, $\mathrm{age}>50\mathrm{Myr}$, and $\mathrm{Z}=[0.004, 0.02]$ (blue); constant SFH, $\mathrm{age}>50\mathrm{Myr}$, $\mathrm{Z}=[0.004, 0.02]$, and assuming energy balance between UV and IR (red); and declining SFH, age free, $\mathrm{Z}=[0.0004, 0.004, 0.02]$, and assuming energy balance (black). Crosses show the synthesised flux in the filters. For each model and galaxy, we show the $\chi^2_r$, the UV $\beta$ slope associated with the best-fit solution, and the predicted infrared luminosity when relevant.}
     \label{fig:sed}
     \end{figure*} 
      \addtocounter{figure}{-1}
       
       \begin{figure*}[htbf]
    \centering
         \includegraphics[width=8.75cm,trim=0cm 0cm 0cm 0cm,clip=true]{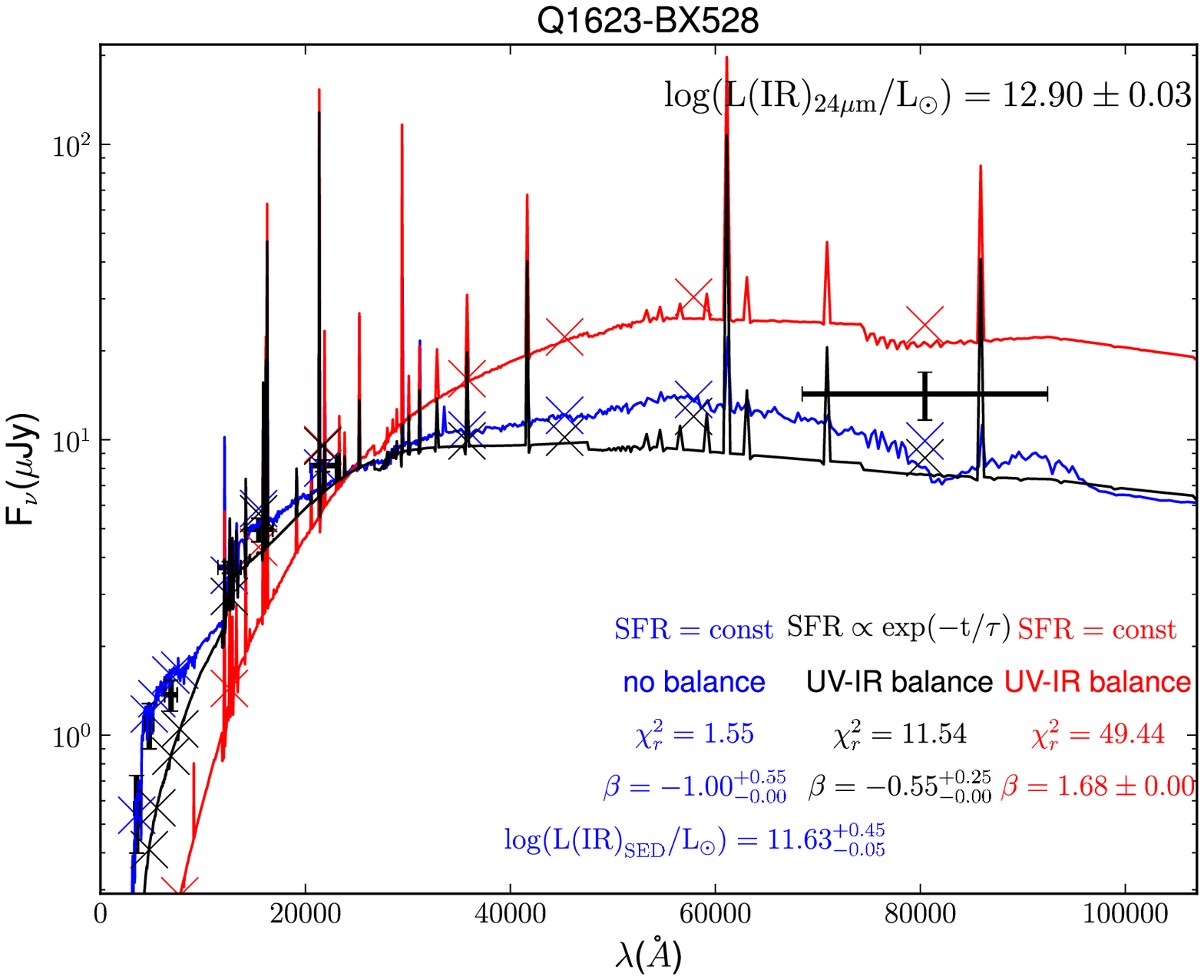}     
     \includegraphics[width=8.75cm,trim=0cm 0cm 0cm 0cm,clip=true]{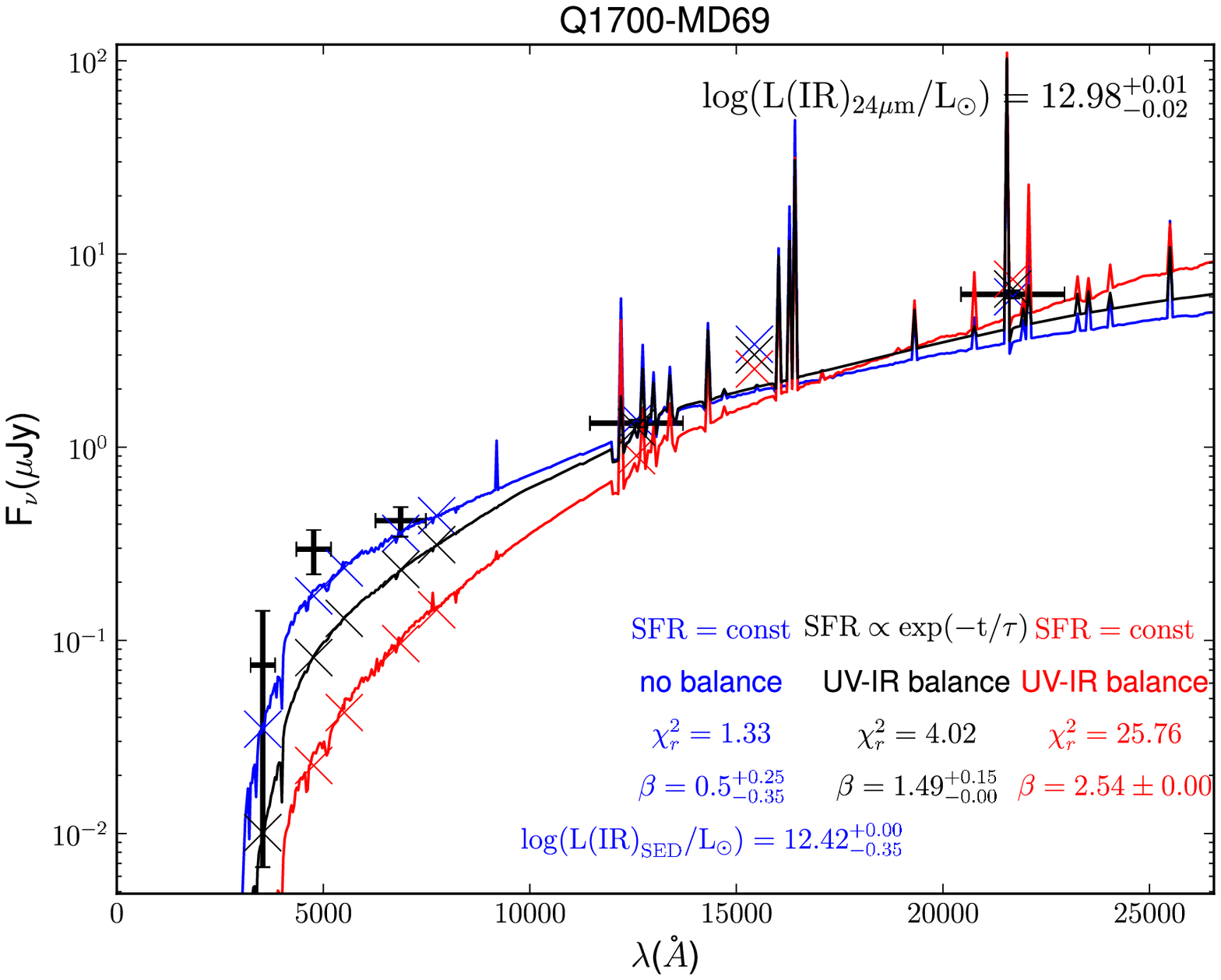}
     \includegraphics[width=8.75cm,trim=0cm 0cm 0cm 0cm,clip=true]{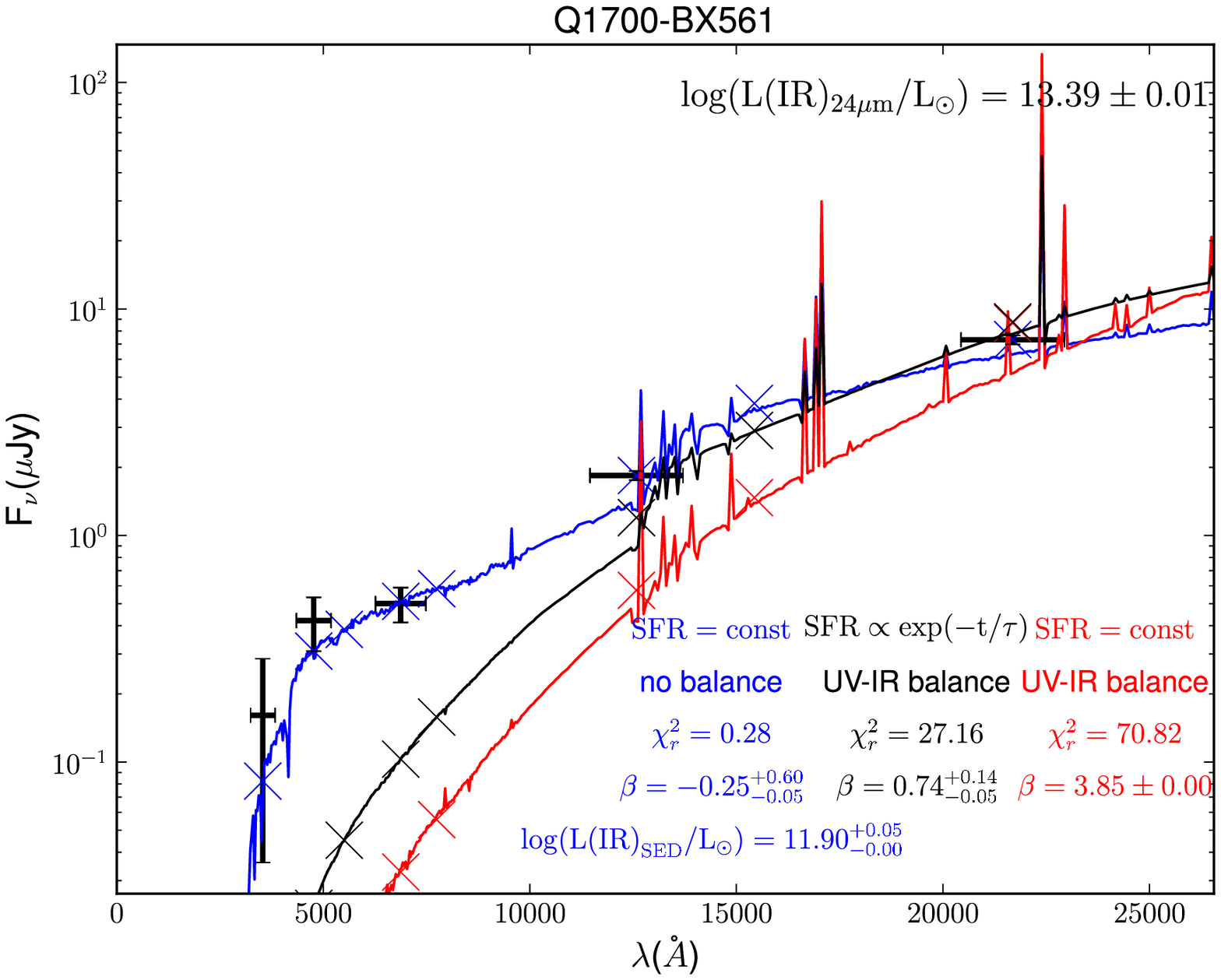}
     \caption{Continued.}
 \end{figure*} 
 
As shown in Figure~\ref{fig:uvir_sed}, for the ULIRGs, 
the SFRs derived from SED fitting are inconsistent with those 
derived from UV and IR luminosity.
The mismatch between
\sfruvir\ and \sfrsed\ arises from the inability of our SED fitting
code to reproduce both the derived IR luminosity and the observed UV
 $\beta$ slope. UV $\beta$ slope can be used to derive the dust attenuation assuming an attenuation curve for typical (i.e., $L^\star$) galaxies at high-redshift \citep[e.g.,][]{reddyetal2010}.
 We show in Figure~\ref{fig:sed} the SED 
fits of the seven ULIRGs in our sample for
three different models: two assuming a constant SFH, with and without
assuming energy balance, and one assuming a declining SFH with relaxed
assumptions on age and metallicity, and assuming energy balance. This latter model is used because it is the model with the maximum number of degrees of freedom and should be more able to reproduce the SED.

The observed $\beta$ slopes are 
bluer that what we would expect 
given the high IR luminosities, assuming a Calzetti curve
\citep{meureretal1999,calzettietal2000}. This is also consistent with
the better agreement between \sfrsed\ and \sfruvir\ for the ULIRGs assuming 
a declining SFH and a large range of ages and metallicities: the
$\beta$ slope is more sensitive to change in age and metallicity for a
declining SFH than for a constant SFH
\citep{leitherer&heckman1995}. Indeed, relaxing assumptions on age and
metallicity for the declining SFH leads to slightly younger ages
($\mathrm{age}>30\mathrm{Myr}$) and SED fits with the minimum allowed
metallicity ($\mathrm{Z}$=0.0004), which 
lead to better reproduce the observed $\beta$ slope, but with
metallicity values unrealistically low \citep[e.g.,][]{kilercieseretal2014}. However,
even this last set of assumptions does not 
yield satisfactory fits of the UV $\beta$ slope for the most IR luminous 
galaxies: the $\beta$ slopes predicted by the SED fits are redder than 
the observed ones.

Our study confirms 
previous results regarding ULIRGs \citep[e.g.,][]{goldaderetal2002,reddyetal2010}: 
 ULIRGs generally have signficantly redder IR/UV luminosity
ratios that would be inferred from their UV slopes assuming the
\citet{meureretal1999, calzettietal2000} attenuation curves.

Furthermore, depending on whether we assume energy balance, we have $\Delta\log\mathrm{f_{\mathrm{EL}}}=\log(\mathrm{f}_\mathrm{EL(obs)}/\mathrm{f}_\mathrm{EL(SED)})$ as high as $\sim1.0$ dex. This is due to the large difference in dust attenuation derivation found for the ULIRGs between the two method used here (with or without energy balance):  the mean difference is $\Delta \mathrm{A}_\mathrm{V}=\mathrm{A}_\mathrm{V,\hspace{1mm} energy\hspace{1mm} balance}-\mathrm{A}_\mathrm{V,\hspace{1mm} no\hspace{1mm} energy\hspace{1mm} balance}=1.6$.

\end{appendices}


\bibliographystyle{apj}
\bibliography{ref}

\end{document}